\documentclass[runningheads]{sn-jnl}
\usepackage[ruled,vlined,linesnumbered]{algorithm2e}
\DontPrintSemicolon
\usepackage[authoryear]{natbib}
\usepackage{float}
\usepackage{caption}
\usepackage{csvsimple}
\usepackage{booktabs}
\usepackage{underscore}
\usepackage{longtable}
\usepackage{graphicx}
\usepackage{subfigure}
\usepackage{amsmath}
\begin{document}

\title[Article Title]{On dynamic multi-agent pathfinding methods: review, simulations and modifications}


\author*[1]{\fnm{Gabriel} \sur{Fejziaj}}\email{gabriel.fejziaj@student.po.edu.pl}

\author[1]{\fnm{Salama} \sur{Hassona}}\email{s.hassona@po.edu.pl}

\author[1]{\fnm{Wieslaw} \sur{Marszalek}}\email{w.marszalek@po.edu.pl}

\affil[1]{\orgdiv{Department of Computer Science}, \orgname{Opole University of Technology}, \orgaddress{\city{Opole}, \postcode{45-758}, \country{Poland}}}


\abstract{This paper presents a systematic study of pathfinding algorithms in the context of Dynamic Multi-Agent Pathfinding (D-MAPF), a setting that combines dynamic obstacles, partial observability, and inter-agent conflicts. We evaluate six representative algorithms—Dijkstra, D* Lite, Space-Time A*, WHCA*, M*, and a novel method denoted as A** within a unified simulation framework. The proposed A** algorithm introduces a template-based approach that decouples offline geometric path generation from online temporal adaptation. By precomputing multiple diverse candidate paths and dynamically reconnecting to them using space-time planning, A** improves solution quality in environments with frequent changes and limited sensing.}

\keywords{Dynamic Pathfinding, Multi-Agent, D-MAPF, Partial obstacle observability, A**}



\maketitle
\section{Introduction}
\label{Introduction}
Path planning the problem of computing a collision-free route for agents moving through various environments, is a foundational task in:
\begin{itemize}
    \item video games (\cite{VideoGames}),
    \item motion of autonomous vehicles (\cite{Cars}),
    \item air traffic control (\cite{Airplanes}),
    \item logistics and supply chain optimization (\cite{Warehouse}),
    \item crowd simulation in urban planning (\cite{Crowd}),
    \item and many others (\cite{MAPF}).
\end{itemize}
Such a task can be found in warehouses, where robots navigate crowded shelves, in real-time strategy (RTS) games, where large numbers of agents must coordinate movements across maps, and when trying to run simulations throughout various environments. The need for efficient and reactive pathfinding spans a wide range of domains. To understand the complexity of modern pathfinding systems the issue needs to be perceived as an evolution of challenges: moving from static, single-agent problems to highly unpredictable, dynamic multi-agent environments.

\subsection{The Evolution of Pathfinding: From Static to Dynamic}

At its most basic level, classical pathfinding assumes a single agent operating in a fully known, static world. The objective of pathfinding systems is to find the shortest sequence of grid cells from a start point to a destination without hitting walls. Algorithms like the classical Dijkstra’s algorithm and A* (\cite{Dijkstra,ASTAR}) solve this optimally by treating the map as a graph and searching for the lowest-cost path.

However, real-world deployments often involve environments that change over time: doors open and close, sensors reveal previously unknown hazards only at a close range, or in simulation scenarios the world model is updated as new data arrives. In video games, this manifests as destructible terrain, teleports or enemy units that block previously clear paths. If we use classical A* algorithm in such a world, the agent must compute a completely new path from scratch every time it encounters a blocked route. This is computationally expensive and impractical for real-time systems.

To bridge this gap, Dynamic Pathfinding was introduced (\cite{DynamicPathfinding}). Instead of throwing away the previous search effort, incremental heuristic search algorithms like D* and D*~Lite (\cite{DSTARLITE}) repair the existing path. They propagate cost corrections only to the portion of the search graph affected by the change. This reduces replanning overhead proportionally to the distance of the change from the agent, allowing smooth, reactive navigation in changing environments.

\subsection{Multi-Agent Pathfinding (MAPF)}

While dynamic pathfinding handles changes in the environment, a different kind of complexity arises when we introduce more agents into the same space. This brings us to the Multi-Agent Path Finding (MAPF) (\cite{MAPF}).

In MAPF, we still assume a static map, but we now must simultaneously compute collision-free paths for a set of agents . The challenge here is a combinatorial explosion. If one agent plans a path ignoring others, they will collide. If we plan for all agents together as a single “super-robot,” the search space grows exponentially with the agent count, making optimal solvers like Conflict-Based Search (CBS) (\cite{CBS}) perform poorly. While CBS-based methods provide optimal solutions, they are not included in this study due to their high computational cost in dynamic and partially observable settings. Practical MAPF systems therefore rely on decoupled approximations like the Cooperative A* (\cite{ASTAR}), where agents plan their motion one by one, treating previous agents’ paths as moving obstacles in a shared reservation table.

\subsection{Dynamic Multi-Agent Pathfinding (D-MAPF)}

The logical consequence of the above is the Dynamic Multi-Agent Pathfinding (D-MAPF) (\cite{D-MAPF}), which is the core concern of this work.

These are the assumptions when experimenting with D-MAPF:
\begin{itemize}
    \item Multiple agents must reach individual goals simultaneously (combatting combinatorial complexity).
    \item The environment is dynamic with obstacles appearing and disappearing on an unknown schedule.
    \item Agents have limited sensors (partial observability) (\cite{PartialObservability}), meaning they only see changes when they are close to an obstacle.
\end{itemize}
Despite its relevance to real-world applications, the literature on D-MAPF with partial observability and prioritized planning (\cite{D-MAPF-PartialObservability}) remains limited. Existing studies typically address one dimension in isolation: reactive single-agent replanning (D*~Lite) or static-world multi-agent optimization (CBS, WHCA*) (\cite{DSTARLITE,CBS,WHCA}).
Our further study is guided by the following research questions:
\begin{enumerate}
    \item 
Which pathfinding algorithm achieves the lowest sum of costs (SoC) (\cite{MAPF}) in a D-MAPF setting with partial observability and dynamic obstacles? \item  What are the differences in  algorithms used in benchmark and in what environments should they be used.
\end{enumerate}
\newpage
\subsection{Contributions}

This paper makes the following contributions:
\begin{itemize}
    \item \textbf{Formalization and Framework:} The D-MAPF setting is formalized as a reactive, partially observable MAPF problem on a grid with time-scheduled dynamic obstacles and per-agent vision ranges. A simulation framework that instantiates this setting for benchmarking is described.
    \item \textbf{Algorithmic Integration:} Six diverse pathfinding algorithms: Dijkstra’s algorithm, D*~Lite, Space-Time A*, WHCA*, M*, \textbf{and the newly proposed A**} are integrated into a common framework, enabling systematic benchmarking and direct performance comparison to determine which algorithm performs best in the defined environment.
    \item \textbf{Vision-Based Obstacle Filter:} A transparent vision filter is introduced that enforces partial observability across all planners without requiring modifications to their core search logic.
    \item \textbf{Empirical Evaluation:} A systematic evaluation is conducted across 8 benchmark maps, 10 agent-count configurations, and 100 repetitions per combination, reporting comprehensive metrics including the SoC, makespan, and failure rates.
    \item \textbf{A** Characterization:} A novel pathfinding algorithm, A**, is introduced as the primary contribution of this work. The method combines offline geometric template generation with online temporal adaptation, enabling efficient replanning in dynamic and partially observable environments. Experimental results show that A** achieves the lowest SoC across most evaluated configurations, demonstrating the effectiveness of the proposed approach in the current setting.
\end{itemize}

\subsection{Problem Constraints and Scope}

To bound the claims made in subsequent sections, the following constraints apply:
\begin{itemize}
    \item \textbf{Grid Topology:} All maps are four-connected grids in MovingAI format. Diagonal movement is out of scope as to reduce computational possibilities and to minimize simulation time.
    \item \textbf{Deterministic Obstacle Schedules:} Dynamic obstacles follow a fixed schedule sampled once at the start from a randomized formula. Agents do not have access to this schedule and must rely on their vision range.
    \item \textbf{Vision Model:} Agents perform ray-casting in the direction of travel up to distance $r$.
\item \textbf{Prioritized planning:} Inter-agent coordination uses a fixed priority order (decoupled approximation), meaning plans are not guaranteed to be globally optimal. 
\item \textbf{Exclusion of CBS:}
Optimal coupled solvers such as Conflict-Based Search (CBS) are not included in this study. While CBS provides optimal solutions in static MAPF settings, its applicability in the considered D-MAPF scenario is limited due to two factors: (i) the high frequency of environment changes (dynamic obstacles appear/disappear), and (ii) partial observability, which causes agents to discover new obstacles incrementally. Both factors may invalidate previously constructed conflict trees, forcing a complete or substantial reconstruction of the search. As a result, CBS-based methods become computationally impractical for the reactive, real-time replanning regime studied in this paper (\cite{CBS,CBSBAD}). There is research on the subject that combines the CBS with D* Lite (\cite{D-CBS}), but because of the amount of work needed to adjust other algorithms this paper is focused on decoupled and incremental methods that are better suited for frequent replanning under dynamic and partially observable conditions and leaves CBS evaluation for future work.
    \item \textbf{Bounded Horizon:} All planners operate with a planning horizon capped at $\min(T_{\text{global}}-t,\; d_{\text{Manhattan}} \times 5 + 80)$ timesteps to prevent state-space explosion (\cite{TimeHorizon}).
    \begin{itemize}
    \item \(T_{\text{global}}\) - maximum allowed duration of the simulation, expressed as the total number of discrete timesteps. This is a fixed constant imposed by the experimental setup (\(T_{\text{global}} = 500\)).
    \item \(t\) - current timestep at the moment the planning procedure is invoked. The agent is situated at time \(t\) and must compute a path forward from this point onward.
    \item \(T_{\text{global}} - t\) - remaining time budget available until the simulation terminates.
    \item \(d_{\text{Manhattan}}\) - Manhattan distance between the agent's current position and its goal location.
    \item \(5\) - heuristic inflation factor. Multiplying the Manhattan distance by 5 provides headroom for necessary waiting actions and detours.
    \item \(80\) - constant safety padding. This additive term guarantees a minimum planning look ahead even when the agent is already very close to its goal. It ensures that the planner can reason sufficiently far into the future to avoid decisions that could lead to deadlocks.
    \item \(\min(\cdot, \cdot)\) - the overall bound is taken as the minimum of the two numbers. This guarantees that the planning horizon never exceeds either the physically remaining simulation time or the scaled-and-padded distance estimate.
\end{itemize}
    The continuous-space planning is not considered in this paper.
\end{itemize}

\subsection{Paper Organization}
The remainder of this paper is organized as follows. Section~\ref{Introduction} (the present section) sets the context, defines the D-MAPF problem, and outlines the contributions of this paper. Section~\ref{Methods} briefly describes six pathfinding algorithms integrated into the framework and explains in detail \textbf{proposed new method denoted as A**}.  Section \ref{RelatedWork} discusses existing algorithms that were not included in the benchmarks but are the groundwork and inspiration for A**.  Section~\ref{Dataset} presents the benchmark maps, dynamic obstacle model, agent parameters, and evaluation metrics. Section~\ref{Results} reports the experimental outcomes across varying numbers of agents. Section~\ref{Discussion} interprets those results analyzing the influence of agent count, and algorithm characteristics. Finally, Section~\ref{Conclusions} summarizes the main findings and suggests directions for future research. Comprehensive tables, statistical analysis and additional visualizations are provided in the Appendix.
\newpage

\section{Methods}
\label{Methods}
\subsection*{Algorithms}
The simulation framework supports six different pathfinding algorithms, each implementing the
same planning interface. They are evaluated under identical conditions: all share the same
map, dynamic obstacle schedule, agent set, and reservation system. The algorithms cover a
range of strategies from classical uniform-cost search to incremental replanning and
cooperative multi-agent planning. This diversity allows us to compare their performance in
terms of solution quality (SoC), computational efficiency, and robustness to
dynamic changes in the environment.

\subsection{Proposal Of A New Algorithm A**}
\begin{figure}[H]
\centering
\includegraphics[width=\textwidth]{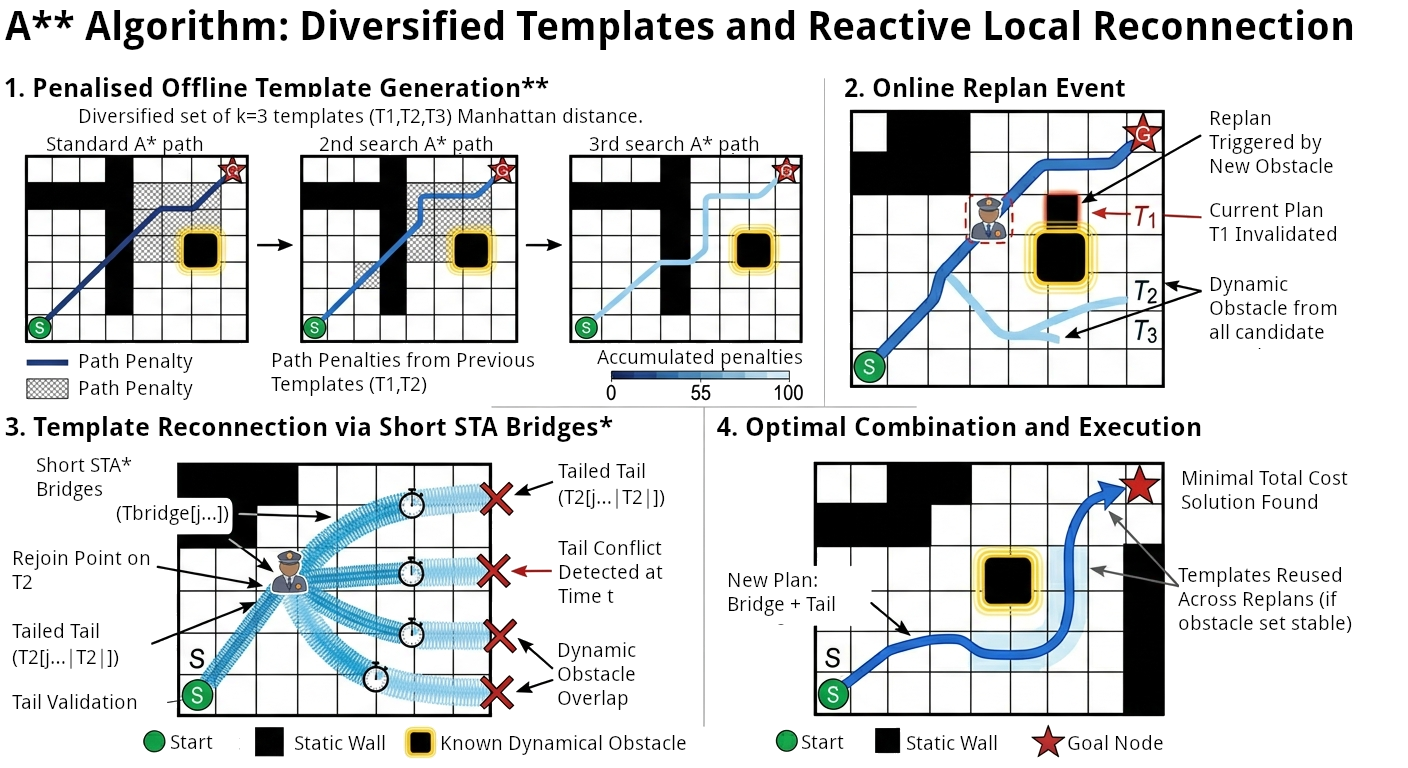}
\caption{A** visualization}
\label{fig:astarstar}
\end{figure}

We designed A** specifically for reactive planning in environments with dynamic obstacles.
The method decouples offline geometric planning from online temporal adaptation through a
template-based representation. During initialization, or whenever the agent's known obstacle
set grows by a newly observed symbol, a set of $k$ candidate paths called \emph{templates}
is generated using a penalized 2D A* search (\cite{ASTAR}). After each template is found, a fixed penalty
is added to the traversal cost of every interior cell along that path, steering subsequent
searches toward geometrically distinct routes and thus diversifying the candidate set. Cells
currently occupied by a known active dynamic obstacle receive an additional penalty, biasing
templates away from hazardous regions without requiring the agent to have full global
knowledge. The parameter $k$ controls the number of candidate paths, (see Algorithm \ref{algo2}). Higher values increase
path diversity at the cost of additional generation overhead. Templates are computed without
temporal constraints and therefore represent lower bounds on achievable path cost in the
absence of inter-agent conflicts.
When a replan is needed, A** attempts to reconnect from the agent's current position to
any accessible point along any stored template. For each candidate the rejoin index $j$ along
template $T_i$, a short STA* (\cite{STA}) segment called a \emph{bridge}, is computed from the current
position to $T_i[j]$, subject to the agent reservations and a per-bridge time limit proportional
to the Manhattan distance. The tail $T_i[j \ldots |T_i|]$ is then validated at the expected
absolute time to ensure that no known dynamic obstacle activates along the remaining
route after the bridge (see Algorithm \ref{algo1}, Fig. \ref{fig:astarstar}). The combination of bridge and tail that minimizes total path length
is selected as the new plan. If reconnection to all existing templates fails, new templates
are generated from the current position and the procedure is repeated. Templates are reused
across replans as long as the agent's known obstacle set has not changed since their
generation, further amortizing the cost of the penalized A* search. The time complexity of \texttt{A**}\ per replanning event is
\(O\bigl(k \cdot |V| \log |V| + k \cdot |P_{\text{max}}| \cdot W \cdot |V| \log |V|\bigr)\),
where \(k\) denotes the template count, \(|V|\) the number of passable cells,
\(|P_{\text{max}}|\) the maximum length among all templates, and \(W\) the
horizon limit applied in Space-Time~A\textsuperscript{*} (STA*).
Template generation requires \(k\) executions of penalized 2D~A\textsuperscript{*}
on the static grid, each bounded by \(O(|V| \log |V|)\), yielding a total of
\(O(k \cdot |V| \log |V|)\).
Bridge construction evaluates up to \(k \cdot |P_{\text{max}}|\) candidate rejoin
points; for each candidate a STA* search is invoked with a horizon of at most
\(W\) time steps, contributing an additional
\(O\bigl(k \cdot |P_{\text{max}}| \cdot W \cdot |V| \log |V|\bigr)\) per replanning event.

\begin{algorithm}[H]
\caption{A** path planning with template regeneration}
\label{algo1}
\KwIn{start $s$, goal $g$, snapshot time $t_{snap}$, start time $t_{start}$, number of templates $k$}
\KwOut{path $P$ or failure}
\If{$s = g$}{
    \Return $\{s\}$\;
}
$cset \leftarrow$ candidate paths for goal $g$\;
\BlankLine
\textbf{// Generate or update templates}
\If{$cset.paths = \emptyset$ \textbf{or} known obstacles changed}{
    $cset.paths \leftarrow$ GenerateCandidates($s,g,t_{snap},k$)\;
    \If{$cset.paths = \emptyset$}{
        \Return failure\;
    }
}
\BlankLine
\textbf{// Try to connect to templates}
$best \leftarrow$ null\;
\ForEach{path $p_i$ in $cset.paths$}{
    \For{$j \leftarrow 0$ \KwTo $|p_i|-1$}{
        $u \leftarrow p_i[j]$\;
        \If{$u$ not passable}{
            continue\;
        }
        \If{$u \neq s$}{
            $bridge \leftarrow$ SpaceTimeA*($s,u$)\;
            \If{$bridge = \emptyset$}{
                continue\;
            }
        }
        \If{tail from $j$ not time-feasible}{
            continue\;
        }
        $combined \leftarrow bridge + tail$\;
        \If{$combined$ valid and shorter}{
            $best \leftarrow combined$\;
        }
    }
}
\If{$best \neq$ null}{
    \Return $best$\;
}
\BlankLine
\textbf{// Fallback: regenerate from current state}
$cset.paths \leftarrow$ GenerateCandidates($s,g,t_{snap},k$)\;
\If{$cset.paths \neq \emptyset$}{
    retry bridge generation\;
}
\Return failure\;
\end{algorithm}

\begin{algorithm}[H]
\caption{GenerateCandidates($s, g, t_{snap}, k$)}
\label{algo2}
\KwIn{start $s$, goal $g$, snapshot time $t_{snap}$, number of paths $k$}
\KwOut{set of candidate paths $P$}
$P \leftarrow \emptyset$\;
$\pi \leftarrow 0$ \tcp*{penalty map}
\For{$i \leftarrow 1$ \KwTo $k$}{
    $p \leftarrow$ 2D A*$(s,g,t_{snap},\pi)$\;
    \If{$p = \emptyset$}{
        break\;
    }
    $P \leftarrow P \cup \{p\}$\;
    \ForEach{cell $x$ in $p$}{
        $\pi[x] \leftarrow \pi[x] + penalty$\;
    }
}
\Return $P$\;
\end{algorithm}

\subsection {Other algorithms}
Detailed algorithmic descriptions (except for our proposed method) are provided in Appendix \ref{A*} . 
Section below presents only high-level summaries of each method, while the full technical details are deferred to the appendix to improve readability.

\subsubsection*{Dijkstra's} algorithm serves as a baseline to evaluate how a static planning method performs in a dynamic environment. It performs a full uniform-cost search at each replanning step and does not use heuristic guidance, which leads to increased computational cost (\cite{Dijkstra}).

\subsubsection*{A*} is not directly evaluated in this benchmark, but serves as a foundation for most of the algorithms used. It combines the path cost from the start with a heuristic estimate of the remaining distance to the goal, prioritizing nodes with the lowest total cost (\cite{ASTAR}).

\subsubsection*{STA*} extends classical A* by incorporating time into the state representation (\cite{STA}). This allows reasoning about collisions between agents. Each state includes both position and timestep, and conflicts are avoided using reservation tables.

\subsubsection*{D* Lite} is an incremental heuristic search algorithm designed for dynamic environments (\cite{DSTARLITE}). Instead of recomputing the entire path after each change, it updates only the affected portions of the existing solution, enabling efficient replanning.

\subsubsection*{M*} is a multi-agent pathfinding algorithm that reduces complexity by planning paths independently and resolving conflicts only when they arise (\cite{MSTAR}). When a collision is detected, the algorithm locally expands the search space to consider joint configurations of the conflicting agents.

\subsubsection*{WHCA*} is a cooperative planning algorithm that restricts planning to a fixed time window (\cite{WHCA}). Each agent plans toward a waypoint within the current window, and the process is repeated until the goal is reached.
\newpage
\section{Similar methods}
\label{RelatedWork}
Several approaches address replanning in dynamic environments through path reuse, multi-path generation, or temporal flexibility, but differ fundamentally from A** in how they combine those aspects.

\textbf{Multi-path replanning} methods, such as the Alternative Paths Planner (APP) (\cite{Inspiration2}) and the Anytime Informed Multi-Path Replanning Strategy (MARS) (\cite{Inspiration3}), maintain multiple candidate trajectories to improve robustness in dynamic settings. The APP precomputes a set of alternative paths offline, enabling a robot to quickly switch routes when an obstacle blocks its primary path (\cite{Inspiration2}). Similarly, the MARS builds a set of precomputed paths and, when a collision occurs, selects the best alternative to resume execution with minimal delay (\cite{Inspiration3}). While these methods share the idea of maintaining path diversity, they operate within a unified planning framework, where geometric path generation and online adaptation are tightly coupled—typically relying on the same underlying planner for both phases. In contrast, A** explicitly decouples geometric path generation (offline templates) from temporal feasibility (online Space-Time A* reconnection). Such separation allows each phase to employ different algorithms and cost functions, offering greater flexibility when switching between precomputed routes during execution.

\textbf{Precomputation with temporal flexibility.} The approach of (\cite{Inspiration1}) focuses on introducing temporal slack into pre-planned multi-agent solutions to absorb disturbances. Their algorithm, FlexSIPP, precomputes all possible plans for a delayed agent by exploiting the temporal flexibility of other agents—the maximum delay an agent can tolerate without forcing a cascade of changes. When a delay occurs, the system quickly retrieves a precomputed adjustment that respects the flexibility bounds of unaffected agents (\cite{Inspiration1}). Unlike that method, A** does not rely on temporal flexibility within a single nominal plan. Instead, it maintains a diverse set of spatial alternatives (templates) generated without any temporal assumptions. When a disruption occurs, A** dynamically reconnects to the most suitable spatial alternative using a time-aware planner (STA*). Thus, while FlexSIPP reasons about \textbf{how much} a plan can be stretched, A** reasons about \textbf{which} spatially distinct route to switch to.

\textbf{Summary of differences.}
A** differs from previously designed algorithms by combining three distinctive elements: (i) it generates explicit path diversity through a penalized template generation that avoids repeatedly producing similar routes; (ii) it performs localized, time‑aware reconnection to stored templates rather than modifying or repairing a single evolving plan; and (iii) it decouples the offline generation of spatial alternatives from the online verification of temporal feasibility, allowing each phase to be optimised independently. None of the discussed methods APP, MARS, or FlexSIPP incorporates all three aspects simultaneously.

\newpage
\section{Dataset}
\label{Dataset}
The experiments are conducted on seven benchmark grid maps from the MovingAI dataset~ (\cite{MAPF}). One map was made to initially test functionality of the framework. To do that it is designed with numerous corridors where dynamical obstacle should trigger bigger reroute. 
Each map represents a two-dimensional grid environment with distinct structural properties. The evaluated maps are defined as follows:\begin{itemize}
    \item \textbf{empty-32-32}, a fully traversable grid without obstacles.
    
    \item \textbf{maze-32}, a map with narrow corridors that constrain movement and force frequent rerouting.
    
    \item \textbf{maze-32-32-4}, a denser and more constrained variant of maze-like environments.
    
    \item \textbf{random-32-32-10} and \textbf{random-32-32-20}, maps with randomly distributed static obstacles, creating irregular navigation patterns.
    
    \item \textbf{room-32-32-4}, \textbf{room-64-64-8}, \textbf{room-64-64-16}, maps composed of rectangular rooms connected by narrow passages, differing in size and number of rooms.
\end{itemize} All maps are originally static. To simulate dynamic environments, they are extended with time-dependent obstacles as described in Section \ref{symbols}.

\subsection{Grid and Dynamic Obstacles}
\label{symbols}

\begin{figure}[H]
\begin{center}
    \subfigure[]
        {\label{original_path}    \includegraphics*[scale=0.6]{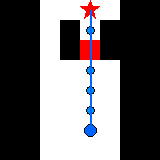}} 
      \subfigure[]
      { \label{changed_path}      \includegraphics*[scale=0.6]{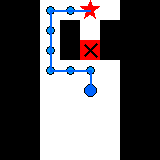}} 
\caption{(a) Original path, (b) Change of path, triggered by a change in status of dynamical obstacles .}
  \label{paths} 
\end{center}
\end{figure}

The environment is a two-dimensional grid of size $H \times W$ loaded from MovingAI \texttt{.map} files.  
Cells are classified as:
\begin{itemize}
    \item \textbf{Passable cells:} denoted by \texttt{.} (empty traversable cell).
    \item \textbf{Static obstacles:} symbols \texttt{@ O T B b} represent permanently blocked cells.
    \item \textbf{Dynamic obstacles:} any symbol other than \texttt{.} and the static obstacle characters is treated as a dynamic obstacle (see Fig. \ref{paths}). Each dynamic obstacle symbol $X$ is associated with a temporal activation rule defined by a schedule.
\end{itemize} The schedule can be deterministic (fixed appear and disappear times) or randomized using uniform distributions.  
For a randomized schedule, the rule is defined as:
\[
X = (t_{\mathrm{start}}^{\min}, t_{\mathrm{start}}^{\max}),\; (t_{\mathrm{end}}^{\min}, t_{\mathrm{end}}^{\max}),\; \tau_{\min},
\]
where $(t_{\mathrm{start}}^{\min}, t_{\mathrm{start}}^{\max})$ specifies the interval from which the appearance time is sampled, $(t_{\mathrm{end}}^{\min}, t_{\mathrm{end}}^{\max})$ the interval for the disappearance time, and $\tau_{\min}$ the minimum required active duration, X is proposed symbol for dynamic obstacle. If the sampled interval would yield a lifetime shorter than $\tau_{\min}$, the activation period is extended to meet this constraint. When the disappearance interval is omitted, the obstacle remains active indefinitely.  
The state of each dynamic obstacle evolves independently, and an obstacle is considered impassable only during its active interval; outside this interval it is passable. For such an approach, the following dynamical obstacles were defined:

\[
X = ({1}^{\min}, 50^{\max}),\; (1^{\min}, 50^{\max}),\; 15^{\min}
\]
\[
Y = ({1}^{\min}, 50^{\max}),\; (1^{\min}, 50^{\max}),\; 15^{\min}
\]
\[
Z = ({1}^{\min}, 50^{\max}),\; (1^{\min}, 50^{\max}),\; 15^{\min}
\]
Thus, the X, Y and Z are defined identically.
\subsection{The Agent Model and Shared Parameters}

Each agent $i$ is defined by a start position $(s_x^i, s_y^i)$, a goal position
$(g_x^i, g_y^i)$ and a vision range $r$ (in cells). The start and goal positions
are generated randomly at the beginning of the simulation using a farthest-first
strategy to ensure sufficient spatial separation.  Let $\mathcal{P}$ be the set
of all passable cells.  The first position is chosen uniformly at random
from $\mathcal{P}$.  Each subsequent start $s_k$ is selected as
\[
s_k = \arg\max_{p \in \mathcal{P} \setminus \{s_1,\dots,s_{k-1}\}}
      \min_{1 \le i < k} \bigl( |p_r - (s_i)_r| + |p_c - (s_i)_c| \bigr),
\]
i.e., the cell whose minimum Manhattan distance to all already placed start positions
is maximized.  Goals are generated independently on $\mathcal{P}$ after
removing all start cells, using the same farthest-first rule.
A permutation of goals is then sought such that for every agent $i$,
\[
|(s_i)_r - (g_{\pi(i)})_r| + |(s_i)_c - (g_{\pi(i)})_c| \;\ge\; L_{\min},
\]
where $L_{\min}$ is the minimum path length (default $6$).  The minimum
distance between any two starting positions (and any two goals) is enforced by the
parameter \texttt{min\_agent\_distance} (default $4$).  All random draws are
controlled by a configurable seed to guarantee reproducibility.

At each timestep, the agent casts a ray along its direction of movement and
records all cells up to distance $r$, stopping at the first static or active
dynamic obstacle.  The agent also knows of all cells
neighboring its current position.  The symbols of any dynamic obstacles
encountered during the entire simulation are stored per agent.  In subsequent
planning, only obstacles whose symbols have already been observed are treated
as impassable; unknown symbols are assumed passable.

\subsection{Reservation System and Planning Order}

To prevent inter‑agent collisions, a shared reservation table is maintained. For each agent that has already planned its path, the table records: \textbf{vertex reservations $(x, y, t)$}, and \textbf{ edge reservations $(x_1, y_1, x_2, y_2, t)$.}
When a new agent plans its path, it must avoid all reservations made by previously planned agents. The planning order is fixed by the agent index given in the input file. This prioritized scheme guarantees that, for a given agent index, the set of constraints is identical regardless of the planning algorithm used.

\subsection{Replanning and Horizon Limits}

All planners operate in a reactive replanning mode. Initially, a plan is generated for every agent in order. During execution, an agent replans from its current position if it encounters a previously unknown active dynamic obstacle or if its current cell becomes occupied by a dynamic obstacle. Replanning uses the latest environment snapshot and the current reservation table.

\subsection{Evaluation Metrics}

The primary performance metric is the sum of costs: 
\[
\text{SoC} = \sum_{i=1}^{n} \bigl(|\text{path}_i| - 1\bigr),
\]
where \(|\text{path}_i|\) is the number of timesteps agent \(i\) takes to reach its goal (including waiting steps).

Secondary metrics are defined as follows:

\begin{enumerate}
    \item \textbf{Makespan} (\cite{MAPF}) – the timestep when the last agent reaches its goal: 
    \[
    \text{Makespan} = \max_{i=1..n} \bigl(|\text{path}_i| - 1\bigr).
    \]

    \item \textbf{Average computational time} (\cite{Methrics}) – the wall‑clock time spent on initial planning plus all replanning events across all agents:
    \[
    T_{\text{total}} = T_{\text{init}} + \sum_{j=1}^{R} T_{\text{replan}}^{(j)},
    \]
    where \(T_{\text{init}}\) is the time for the initial planning phase, \(R\) is the total number of replanning events, and \(T_{\text{replan}}^{(j)}\) is the wall‑clock time of the \(j\)-th replanning.

    \item \textbf{Number of replanning events} (\cite{Methrics2}) – the total count of replanning triggers across all agents:
    \[
    R = \sum_{i=1}^{n} r_i,
    \]
    with \(r_i\) being the number of times agent \(i\) invoked replanning during the simulation.

    \item \textbf{Average replanning time} (\cite{Methrics}) – the mean wall‑clock time per replan, reported per algorithm:
    \[
    \bar{T}_{\text{replan}} = \frac{1}{R} \sum_{j=1}^{R} T_{\text{replan}}^{(j)},
    \]
    with the convention that if no replanning occurred (\(R = 0\)), the average is defined as zero.
\end{enumerate} All experiments are run with identical map, dynamic obstacle schedule, agent set, and random seed (where applicable), ensuring that differences in SoC and runtime are attributable solely to the planning algorithm.

\newpage
\section{Results}
\label{Results}
\begin{figure}[H]
\begin{center}
    \subfigure[]
        {\label{original_path}    \includegraphics*[scale=0.105]{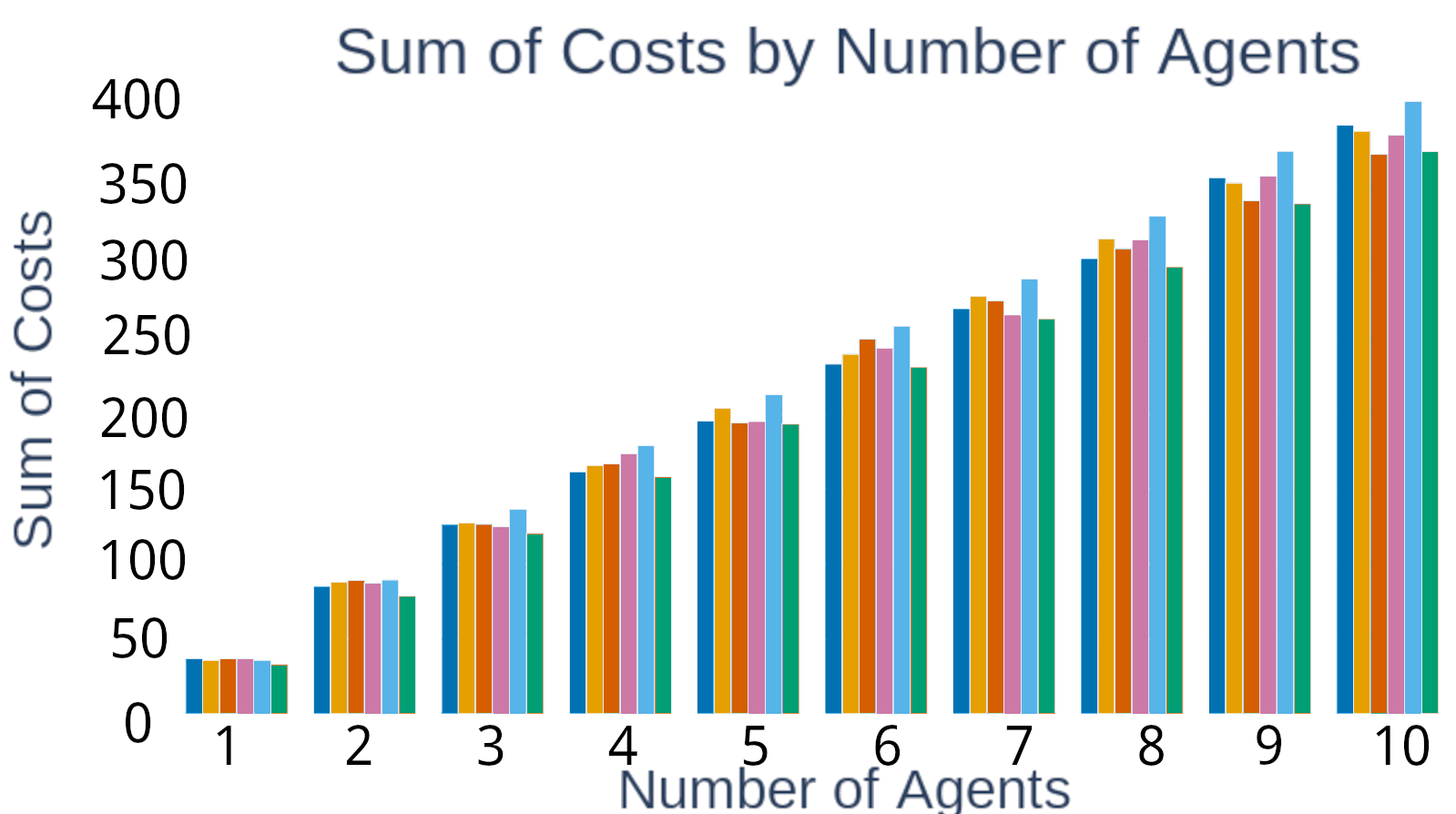}} 
      \subfigure[]
      { \label{changed_path}      \includegraphics*[scale=0.105]{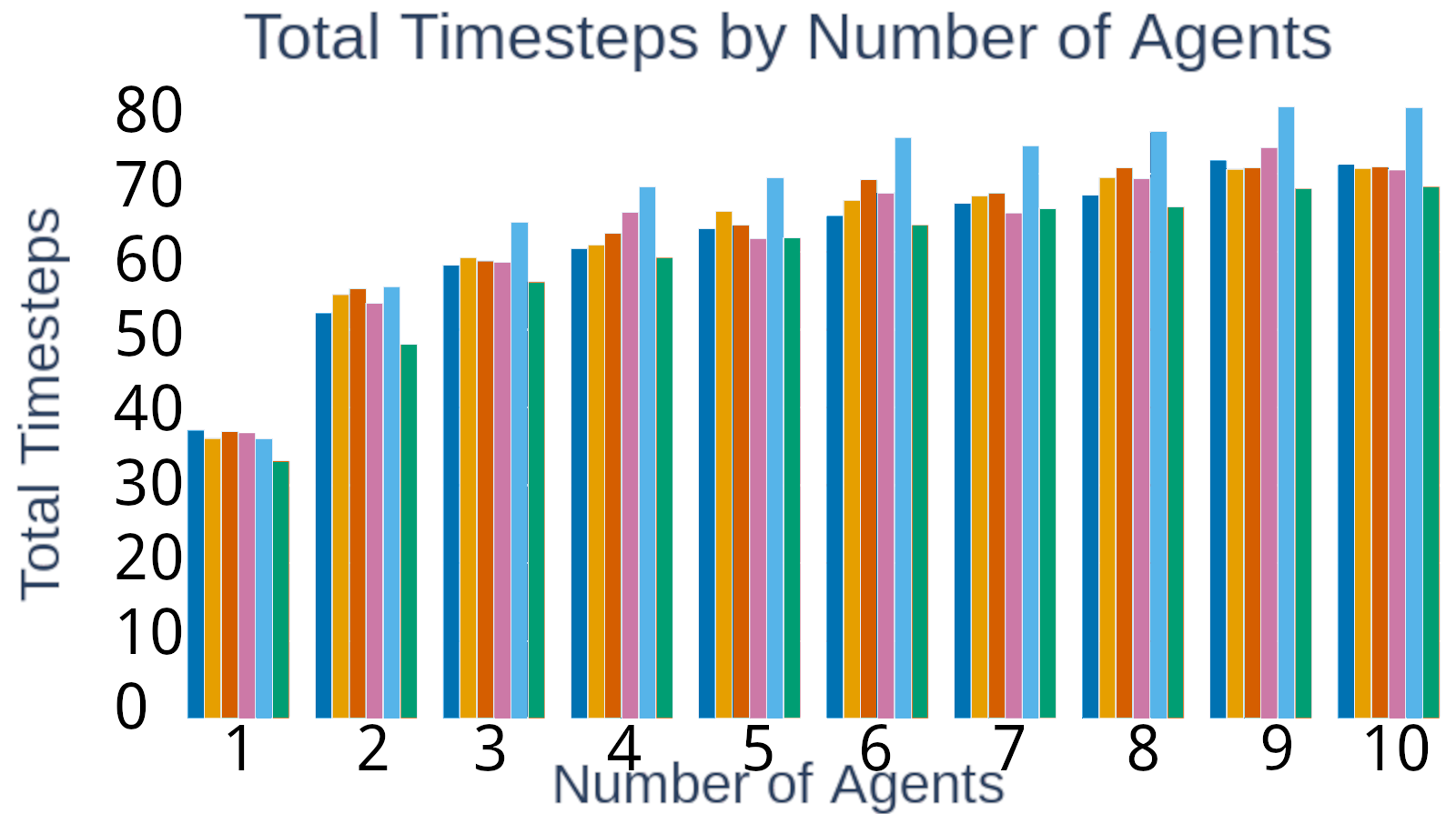}}
      \subfigure[]
      { \label{changed_path}      \includegraphics*[scale=0.105]{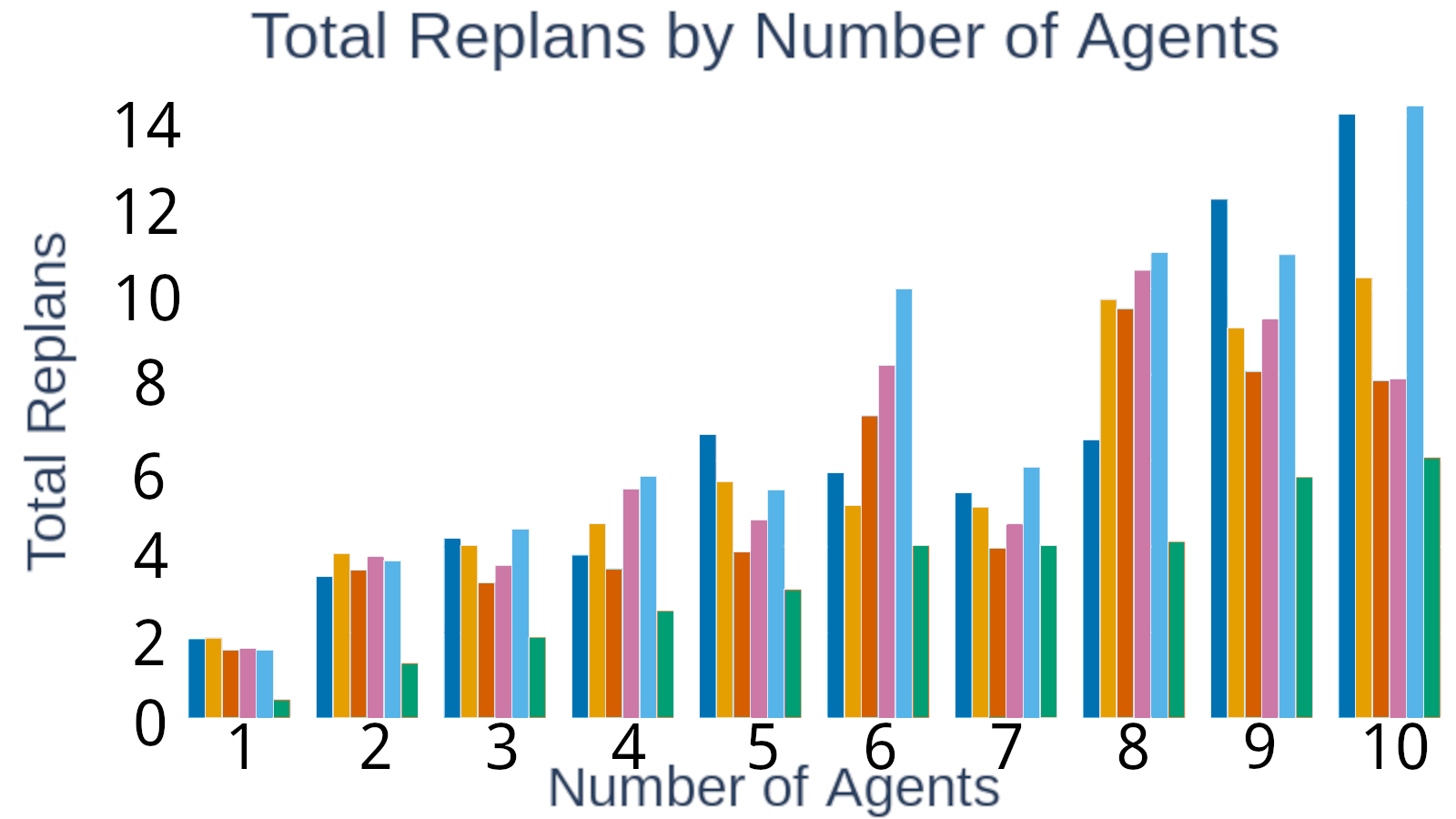}}
            \subfigure[]
      { \label{changed_path}      \includegraphics*[scale=0.105]{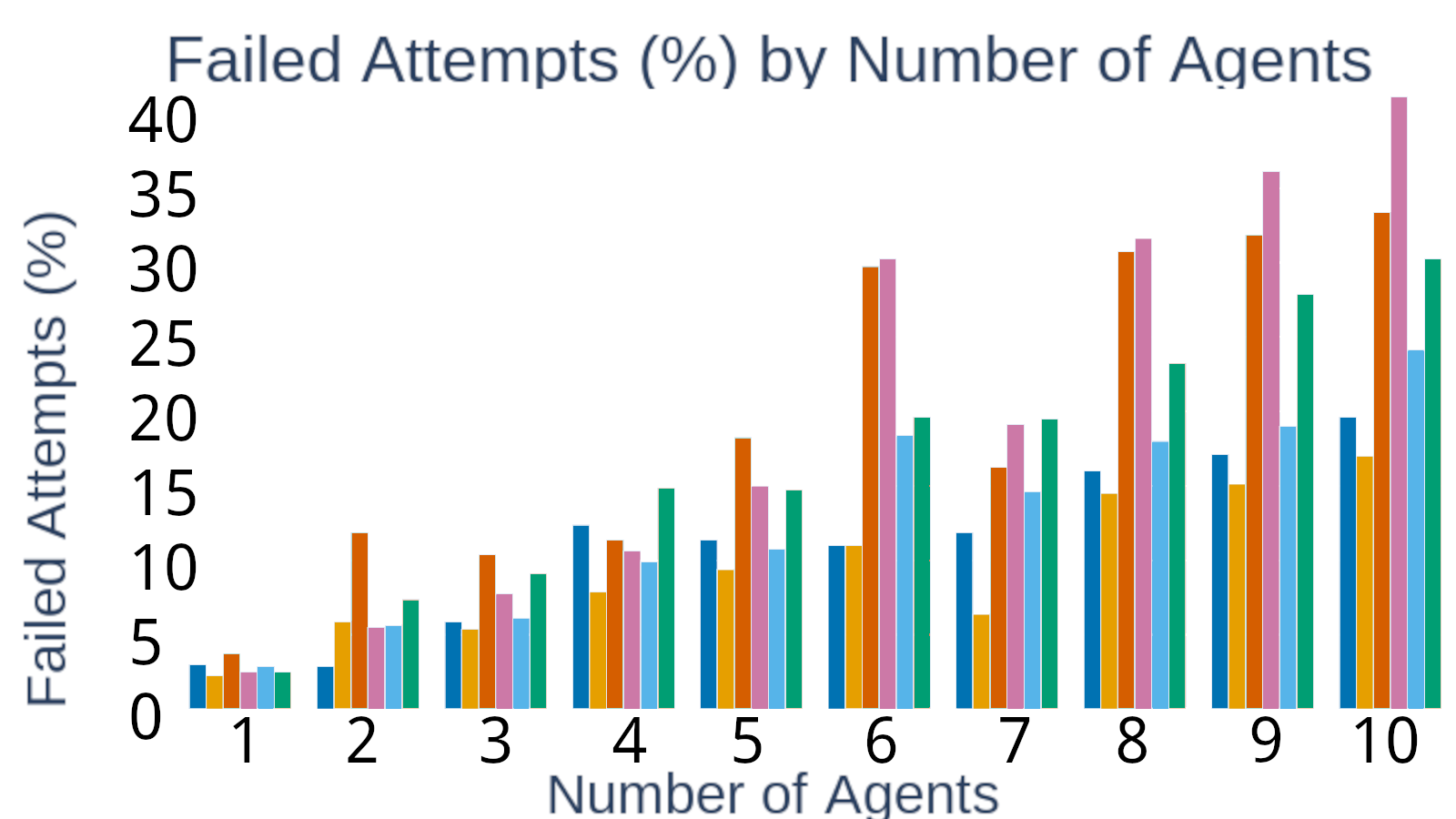}}
       \subfigure[]
        {\label{original_path}    \includegraphics*[scale=0.105]{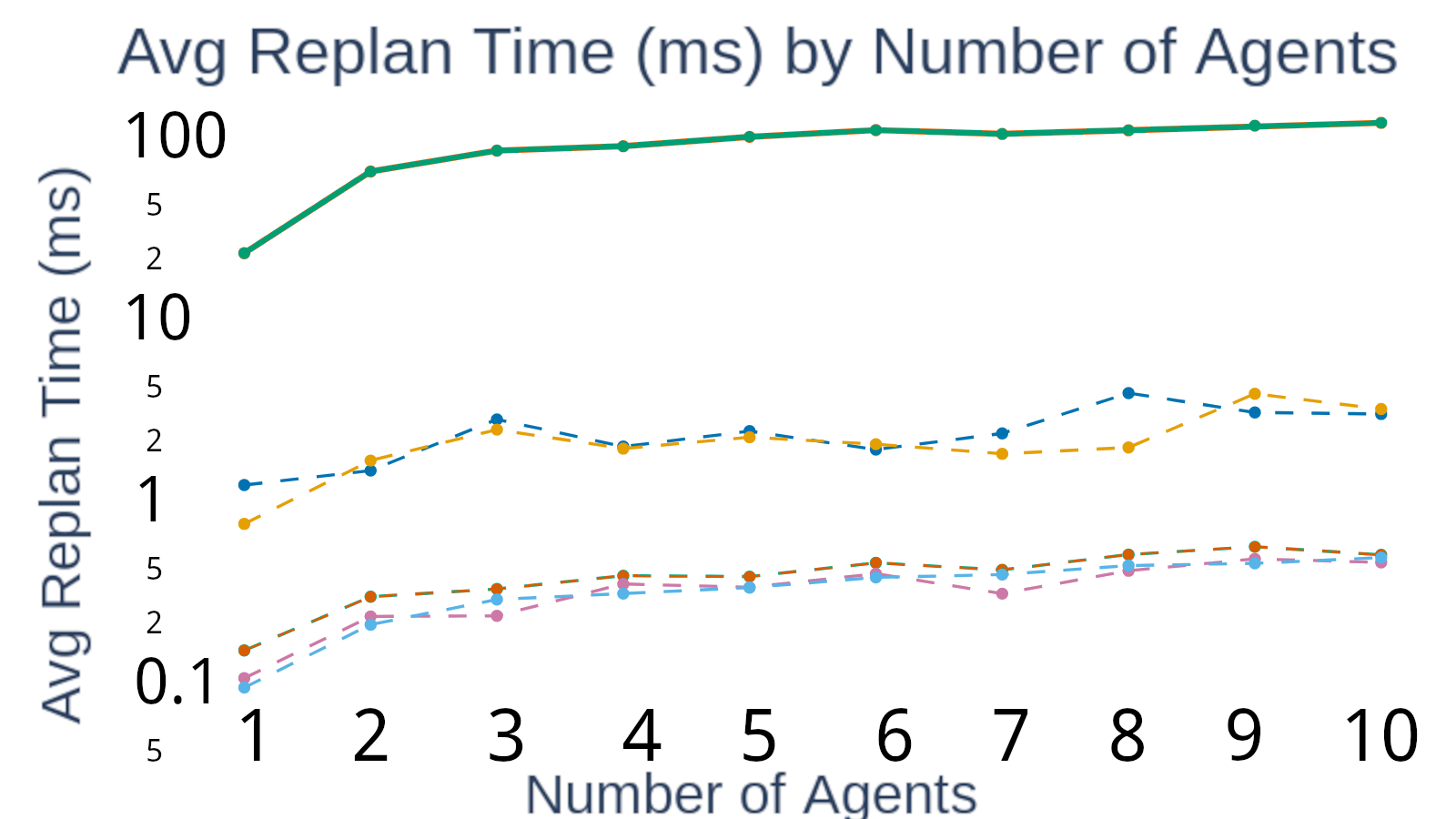}} 
      \subfigure[]
      { \label{changed_path}      \includegraphics*[scale=0.105]{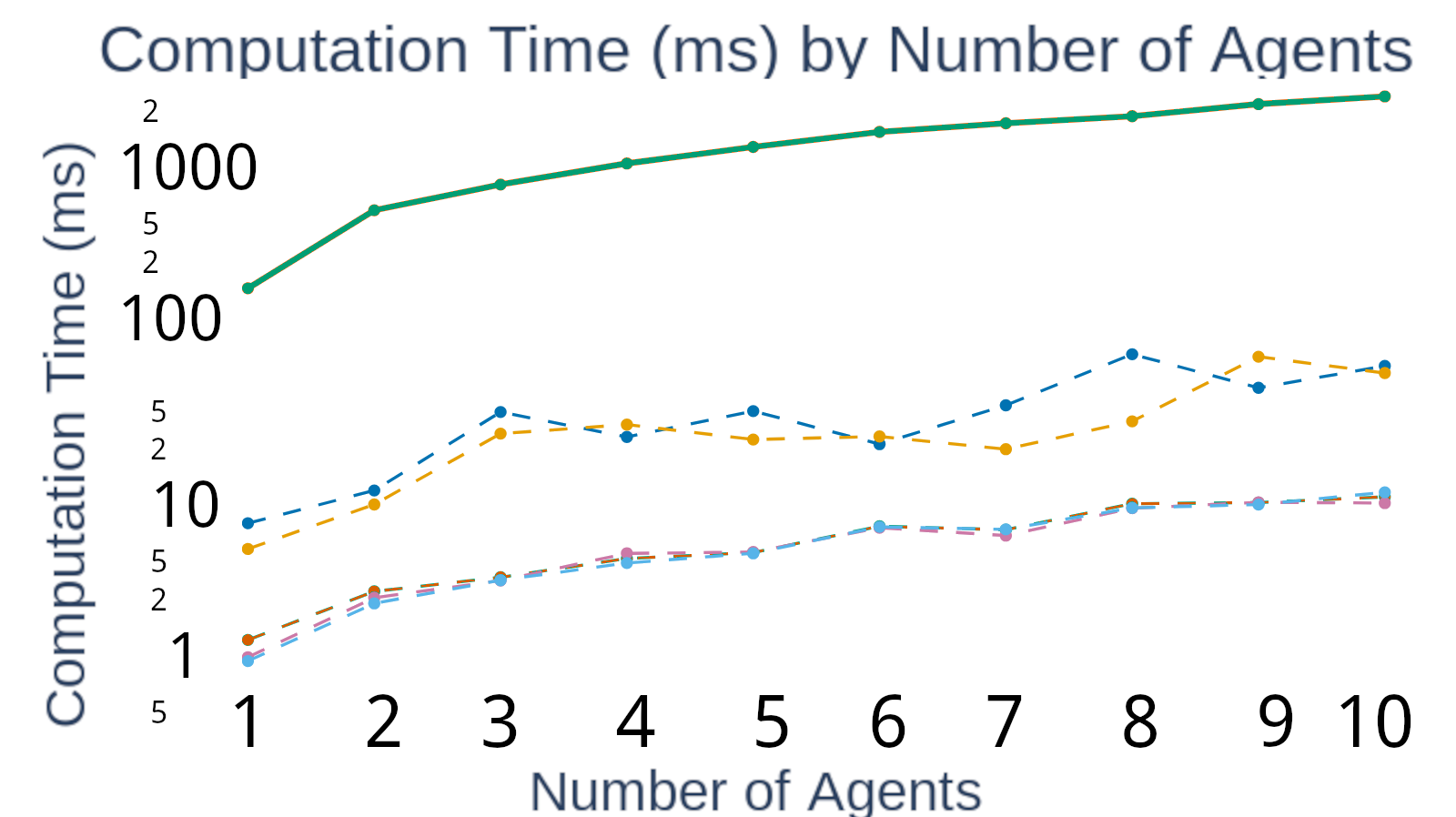}}

      \subfigure[]
      { \label{changed_path}      \includegraphics*[scale=0.8]{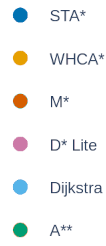}}
      \hspace{5cm}
      \subfigure[]
      { \label{changed_path}      \includegraphics*[scale=0.8]{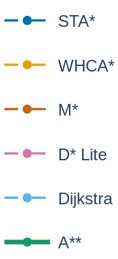}}
      
\caption{Summary of metrics (a) SoC, (b) Makespan, (c) Total replans, (d) Failed attempts, (e) Average replan time, (f) Average computation time, (g) Legend for charts (a)-(b), (h) Legend for charts (e)-(f). In all charts: lesser value at y-axis, better the score.}
  \label{Charts} 
\end{center}
\end{figure}

All six algorithms were evaluated under the same
conditions: map, randomly sampled dynamic obstacle schedule
(obstacles X, Y, Z each with appearance interval $[1, 50]$, disappearance interval
$[1, 50]$, and minimum active duration $\tau_{\min} = 15$), agent placement,
and random seed. Fig. \ref{Charts}. summarizes
the aggregated metrics across all maps. The averaged per-algorithm results show dominance of A ** method across SoC and Makespan. The worst metric for A** is Average replanning. It reaches 16661 ms at its worst configuration. Because of that the average computational time rises drastically up to 1552.44 ms. All benchmarks were run sequentially on a single machine. Specifications are as follows:
    \begin{itemize}
        \item CPU: AMD Ryzen 5 9600X,
        \item GPU: AMD Radeon RX 6700 XT,
        \item RAM: 32GB CL30 6000MT/s DDR5,
        \item OS: Artix Linux.
    \end{itemize}Detailed results are presented in Appendix \ref{Tables}.

\newpage    

\newpage
\section{Discussion of Results}
\label{Discussion}
The experimental results reported in the Tables 1-7 in this paper and the GitHub (see
Appendix \ref{Tables}) reveal systematic performance differences among the six evaluated algorithms across increasing agent counts (1 to 10) and eight benchmark maps. The discussion focuses on three key dimensions: solution quality measured by SoC, computational efficiency (total computation time and replanning overhead), and robustness (failure rate).

\subsection{Solution Quality (SoC)}

A\textbf{**} consistently achieves the lowest SoC for all agent counts from 1 to 9. The only exception is the case of 10 agents, where M* yields a slightly lower SoC (373.07 vs. 375.03 for the best A** variant). D* Lite beats A** in Makespan at 5 and 7 agents, but A** remains superior in terms of SoC for those agent counts. This result stems from the template mechanism: by pre‑computing multiple geometrically distinct global routes A** can bypass regions that become temporarily blocked. The penalty system further encourages templates to avoid known dynamic obstacles, reducing the need for costly online detours.

Dijkstra exhibits the highest SoC amongst all agent counts except for n=1. This is due to the lack of a time dimension in its core search: it treats reservations as static blocked cells within a short horizon without the ability to wait. This forces agents to wait more often or take longer detours, inflating the total cost.

\subsection{Computation Time and Replanning Overhead}

Dijkstra and D* Lite are the fastest algorithms, with total computation times below 6 ms on average. Their simplicity keep the search space small. D* Lite’s incremental nature provides additional speed when changes are local.

STA* and WHCA* have moderate computation times. WHCA* is consistently faster, even 2 times more than STA*, because it limits the STA* search to a window of $W=16$ steps.

A\textbf{**} exhibits significantly higher computation times, ranging from hundreds to thousands of milliseconds. The cost is dominated by two factors: (1) generating $k$ templates requires $k$ penalised 2D A* runs, and (2) bridges attempt to evaluate every cell of all templates, each requiring a full STA* search. The penalty mechanism increases diversity of templates, but brings significant computational overhead.

Replan counts follow a different pattern. WHCA* and Dijkstra replan more frequently, while A** and M* less often. This is the consequence of the way these algorithms work. WHCA* only produces paths with collisions in mind, up to a certain waypoint while the Dijkstra algorithm on the other hand completely ignores time reservations. M* replans only in cases when the dynamical obstacle comes in path, if it comes into contact with other agent algorithm either waits, or backs away. A **  templates anticipate many future conflicts, reducing the need for reactive changes. However, each replan is extremely expensive, so the product “replan frequency × replan time” is still much higher for A**.
\subsection{Robustness and Failure Rate}

The failure rate quantifies the proportion of simulation runs that do not terminate successfully. A run is considered a failure if any agent fails to reach its goal before the global time step limit \(T_{\text{max}}\) (set to 500 in our experiments). Such failures can arise from several causes: (i) the planner may declare that no path exists within the planning horizon (e.g., when all neighbors are reserved by other agents), (ii) the agent may enter a deadlock from which it cannot escape, (iii) the planning algorithm may time out during a replanning attempt. In all these cases, the simulation stops without reaching a complete solution. The failure rate is therefore a critical metric for practical deployment, as it directly reflects an algorithm's ability to handle congested and dynamic environments reliably.

\begin{table}[h]
\centering
\caption{Average failure rate}
\label{tab:failure_rates_sorted}
\begin{tabular}{lc}
\toprule
\textbf{Algorithm} & \textbf{Fail Percentage} \\
\midrule
$\mathit{WHCA}^*$             & 9\% \\
$\mathit{STA}^*$   & 11\% \\
$\mathit{Dijkstra}$           & 13\% \\
$\mathit{D}^*\mathit{lite}$ & 20\% \\
$M^*$                       & 20\%  \\
$\mathit{A}^{**} (Averaged)$   & 21\% \\
\bottomrule
\end{tabular}
\end{table}
    STA*, WHCA* have the lowest failure rates (below 11\%). Their exhaustive time‑expanded search provides strong empirical robustness and is more likely to find a solution within the planning horizon compared to other evaluated methods.
    Dijkstra’s algorithm yields an average failure rate of 13\%. This moderate average conceals a strong dependence on agent density: the failure rate remains below 6\% for up to three agents but rises sharply to over 15\% when 6 or more agents are present (see Appendix \ref{apendixc}).
    D* Lite, A** and M* show the highest failure rates around 20\%, because the first two may temporarily declare a path impossible when reservations block all neighbours, even though waiting a few steps would resolve the conflict. The third however fails as the policy can lead the agent into a dead end of a corridor, from which no policy‑following escape exists.

\newpage
\section{Conclusions}
\label{Conclusions}

A systematic evaluation of six pathfinding algorithms (Dijkstra, D* Lite, STA*, WHCA*, M* and A**) has been performed in a D-MAPF setting that includes dynamic obstacles, partial observability, and inter-agent conflicts. The experiments were conducted on seven MovingAI benchmark maps and one custom‑designed map, with agent counts ranging from 1 to 10.
\textbf{The conclusions are specific to the evaluated setting and may not generalize to all D-MAPF scenarios.}
Based on the obtained results (see Appendix \ref{apendixc}), the following conclusions can be drawn:

\begin{enumerate}
        \item \textbf{A** achieves the lowest SoC at most, but not all, agent
    levels.}
   A** produced the best average SoC for all agent counts from 1 to 9, with improvements over the next best baseline ranging from 0.4\% to 8.5\% (see Table \ref{tab:agents1} and \ref{tab:agents10}). The only exception is n = 10, where M* achieves the lowest SoC  (0.53\% better than the best A** variant (see Table \ref{tab:agents10})). When considering the highest to lowest SoC range, the biggest improvement can be observed at $n=2$ (13.58\%). This result confirms that the template mechanism is most effective when reservation tables are not dense, meaning not many agents in simulation. The improvement comes at a computational cost much higher than any baseline method. A** therefore is to be used in environments where quick readjustments are not crucial, and computing resources are not limited.

    \item \textbf{WHCA* is the most robust algorithm in the evaluated setting.}
Its sliding‑window cooperative planning (window size 16) reduces failure-rate, while incurring only a minor increase in SoC. Its average fail rate is 9\%  – the lowest among all evaluated algorithms – and it never exceeds 17\% even with 10 agents (see Table \ref{tab:agents10}), which produces the highest failure rates opposed to other instances. Second to WHCA* is STA* with failure rate at 11\% (see Table \ref{tab:failure_rates_sorted}) .WHCA* also maintains moderate computation times (e.g. 42 ms at 10 agents), offering the best trade‑off between quality and stability. WHCA* is therefore a strong candidate for a default planner in large‑scale simulations where stability is crucial.

    \item \textbf{Dijkstra's algorithm and D*~Lite are the fastest in computation.}
     Both algorithms complete planning fastest at all agent levels (from 1 to 6 ms). Their simplicity keep the search space small. However, Dijkstra produces the highest SoC among all algorithms for agent count $>$ 1 (see Tables \ref{tab:agents1} and \ref{tab:agents10}). That is because Dijkstra is not a dynamical planner and rising agent counts add more dynamic decisions that pathfinding algorithm must make in a priority-planed simulation.
    
    \item \textbf{M* provides competitive solution quality.}
    M* achieves the best SoC among existing algorithms at $n = 5$, $n = 9$, and
    $n = 10$ (see Table \ref{tab:agents10}), owing to its policy-guided subdimensional expansion that restricts search
    to individually optimal directions.

\end{enumerate}
\newpage
\section*{}
\subsection*{\textbf{Author contributions}} 
Conceptualization, G.F.; methodology, G.F., S.H.; software, G.F.;
validation, S.H., W.M.; formal analysis, G.F., S.H. and W.M.; resources, G.F., S.H. and W.M.; data curation, G.F.; writing—original draft preparation, G.F., S.H. and W.M.; visualization, G.F.; supervision, S.H.; project administration, W.M. All authors have read and agreed to the submitted version of the manuscript. 
 
\subsection*{\textbf{Funding}}
This research did not receive funding.
 
\subsection*{\textbf{Data availability}}
Data analyzed in this research is included in the paper.

\section*{\textbf{Declarations}}
\subsection*{\textbf{Conflict of interest}}
The authors declare that they have no conflict of interest

\appendix
\section{Algorithms}
\subsubsection*{Dijkstra's Algorithm}
\label{Dijkstra}
\begin{figure}[H]
\centering
\includegraphics[width=\textwidth]{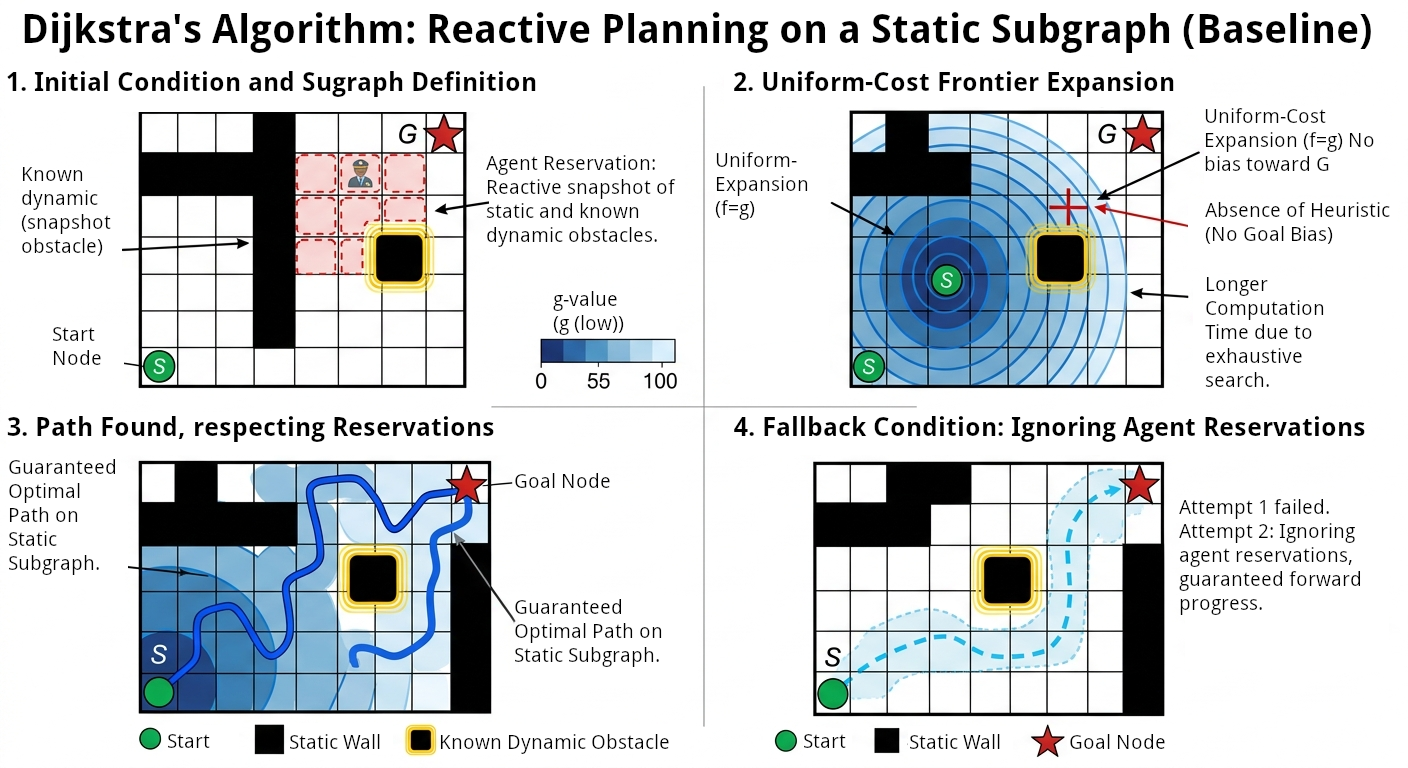}
\caption{Dijkstra's algorithm visualization}
\label{fig:dijkstra}
\end{figure}

Dijkstra's algorithm serves as a reactive planning baseline. At each replanning event, a
static snapshot of the current obstacle state combining static walls, known dynamic
obstacles, and a bounded-horizon agent reservation table is materialized into a
blocked-cell set. Uniform-cost search is then executed on this static subgraph from the
agent's current position to its goal. The absence of a heuristic guarantees correctness
regardless of the structure of the obstacle map, but results in longer computation times
compared to A*-based methods, particularly on large maps. If no path is found subject to
the reservation constraints, a second attempt is made ignoring agent reservations, accepting
potential conflicts in exchange for guaranteed forward progress. Visualization of algorithm is presented by Fig. \ref{fig:dijkstra}. (\cite{Dijkstra}).

\subsubsection*{A*}
\label{A*}
\begin{figure}[H]
\centering
\includegraphics[width=\textwidth]{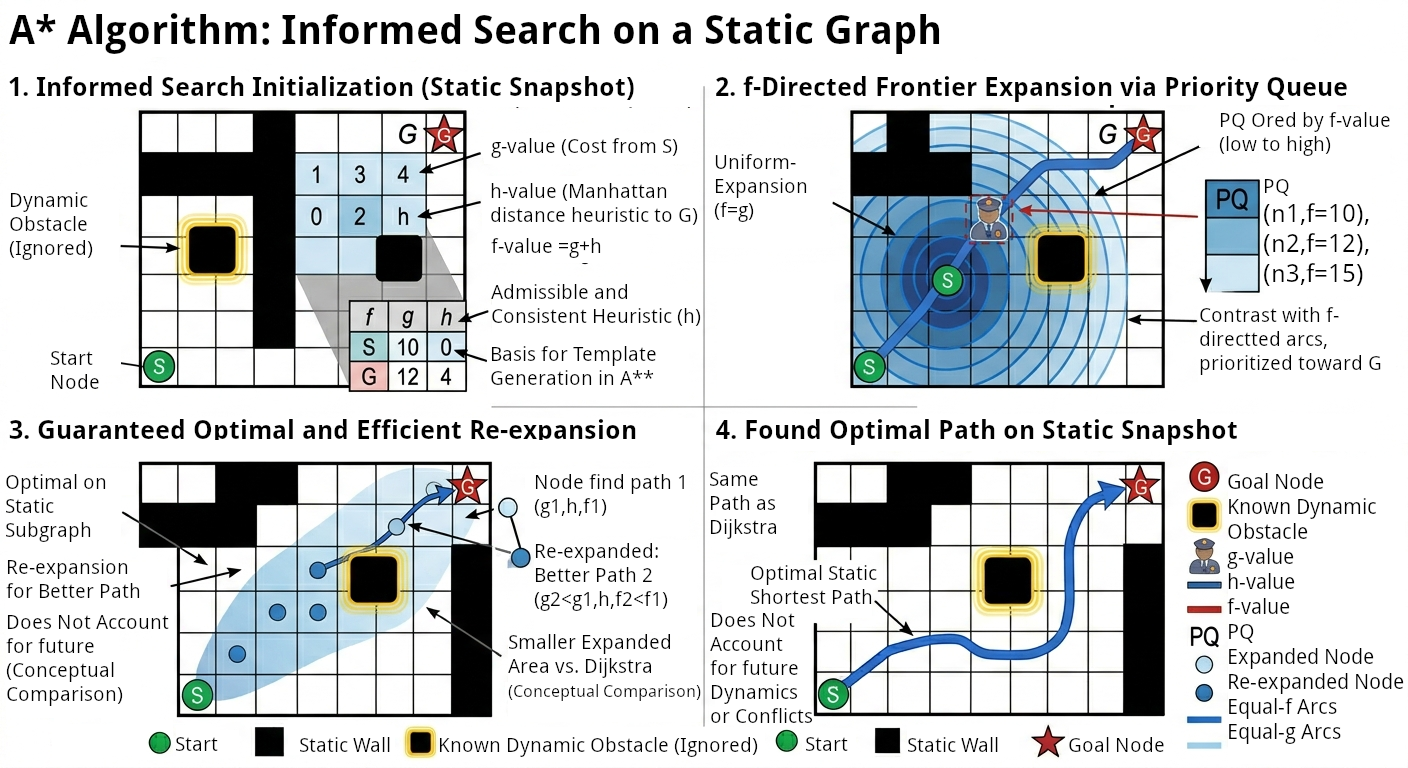}
\caption{A* algorithm visualization}
\label{fig:astar}
\end{figure}

The classical A* algorithm computes the shortest path between two positions in a weighted
graph by combining the actual cost from the start ($g$-value) with an admissible heuristic
estimate of the remaining cost to the goal ($h$-value). In this work the Manhattan distance
serves as the heuristic, which is both admissible and consistent on a four-connected grid.
It maintains a priority queue ordered by $f = g + h$ and expands nodes in non-decreasing
order of $f$, guaranteeing that the first path found to the goal is optimal. The algorithm
operates on a static, two-dimensional snapshot of the environment and does not account for
time or inter-agent conflicts. A* is being used in generation of A** templates. Visualization of algorithm is presented by Fig. \ref{fig:astar}. (\cite{ASTAR}) .

\subsubsection*{Space-Time A* (STA*)}
\label{STA*}
\begin{figure}[H]
\centering
\includegraphics[width=\textwidth]{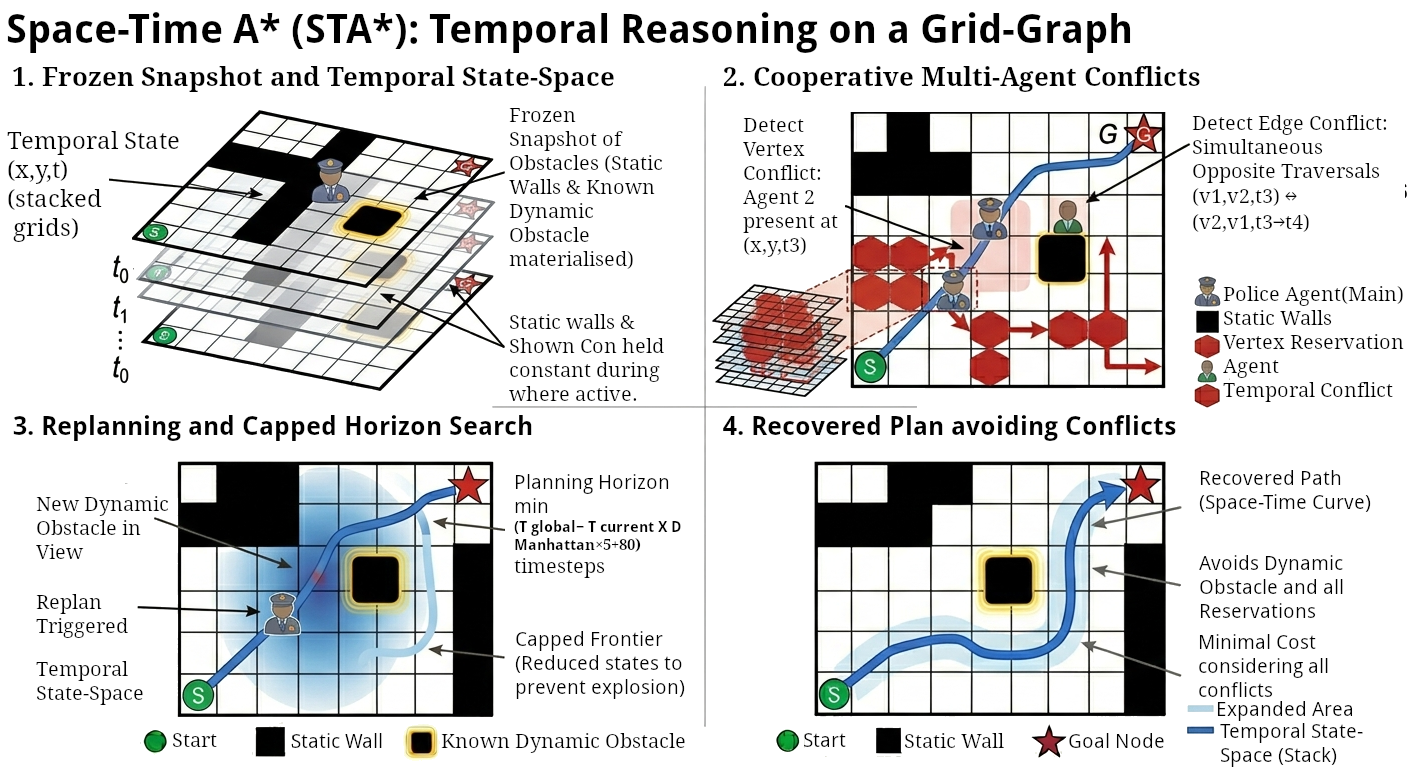}
\caption{STA* visualization}
\label{fig:sta}
\end{figure}

Space-Time A* extends A* by introducing time as a third dimension of the search space.
Each state is represented as a $(\textit{position}, \textit{timestep})$ pair, enabling the
planner to reason about temporal conflicts that arise from the simultaneous movement of
multiple agents. Obstacle states are evaluated at the moment of replanning and held constant
throughout the search a frozen-snapshot model that reflects the agent's partial
observability of the environment. Conflicts with other agents are handled through two
reservation tables: a vertex reservation table that marks positions occupied by other agents
at specific absolute timesteps, and an edge reservation table that prevents two agents from
traversing the same edge in opposite directions within the same step. When a dynamic obstacle
enters the agent's field of view a full replan is triggered from the current position. To
prevent state-space explosion on large maps, the planning horizon is capped at
$\min\!\left(T_{\textit{global}} - t_{\textit{current}},\;
d_{\textit{Manhattan}} \times 5 + 80\right)$
timesteps, where $d_{\textit{Manhattan}}$ is the Manhattan distance from the current
position to the goal. Visualization of algorithm is presented by Fig. \ref{fig:sta}. (\cite{STA}).

\subsubsection*{D* Lite}
\label{D*Lite}
\begin{figure}[H]
\centering
\includegraphics[width=\textwidth]{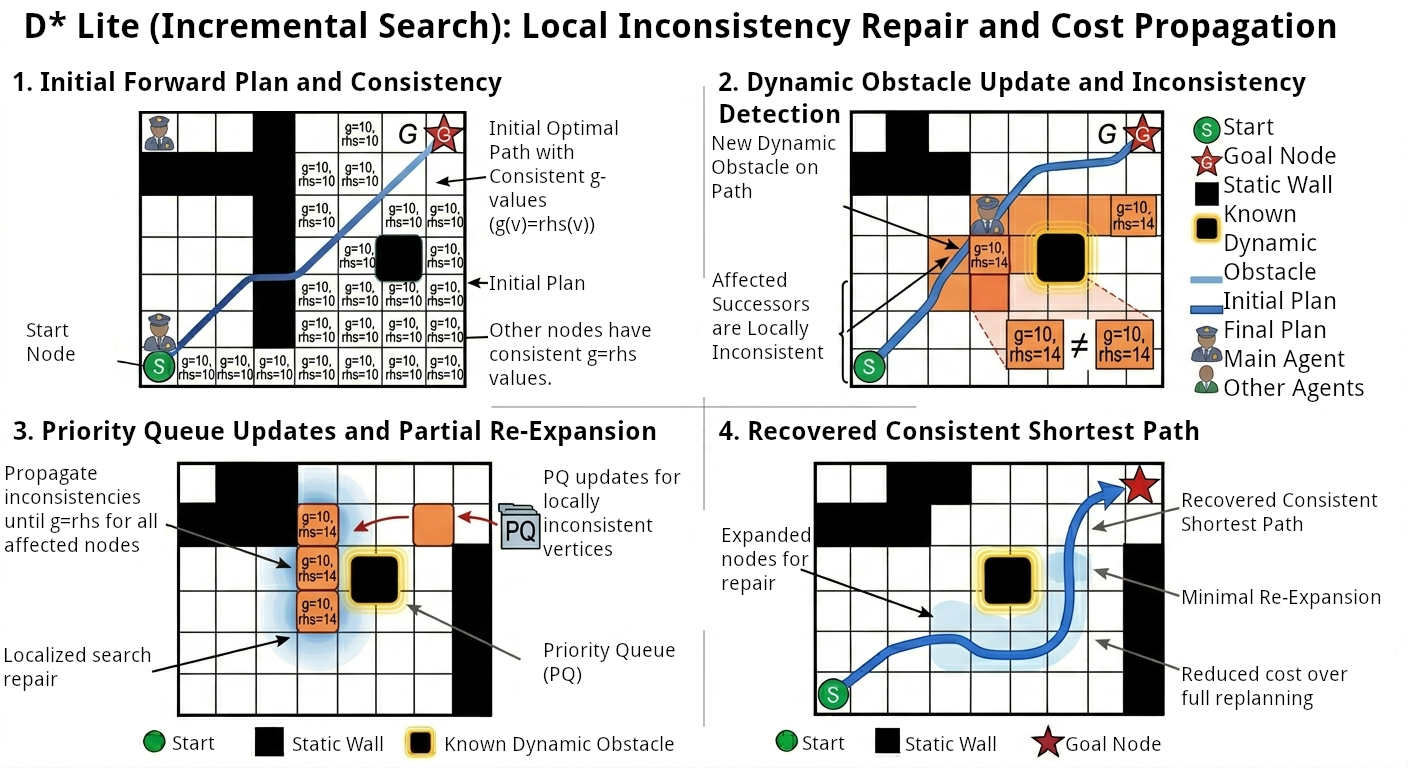}
\caption{D* Lite visualization}
\label{fig:dlite}
\end{figure}

D* Lite is an incremental heuristic search algorithm designed for replanning in partially
known environments. Rather than recomputing the shortest path from scratch after each
obstacle update, it maintains the previous search graph and repairs only the portions
affected by changes. The key quantity is \emph{local inconsistency}: a vertex is locally
inconsistent when its $g$-value deviates from its look-ahead value, defined as the minimum
cost achievable through its successors. Only inconsistent vertices are re-expanded, yielding
computational savings proportional to the number of changed edges. When new dynamic
obstacles are observed, the affected cells are injected into the priority queue and cost
updates are propagated until all inconsistencies are resolved, and a consistent shortest path
is recovered. Visualization of algorithm is presented by Fig. \ref{fig:dlite}. (\cite{DSTARLITE}).

\subsubsection*{M* (Subdimensional Expansion)}
\label{M*}
\begin{figure}[H]
\centering
\includegraphics[width=\textwidth]{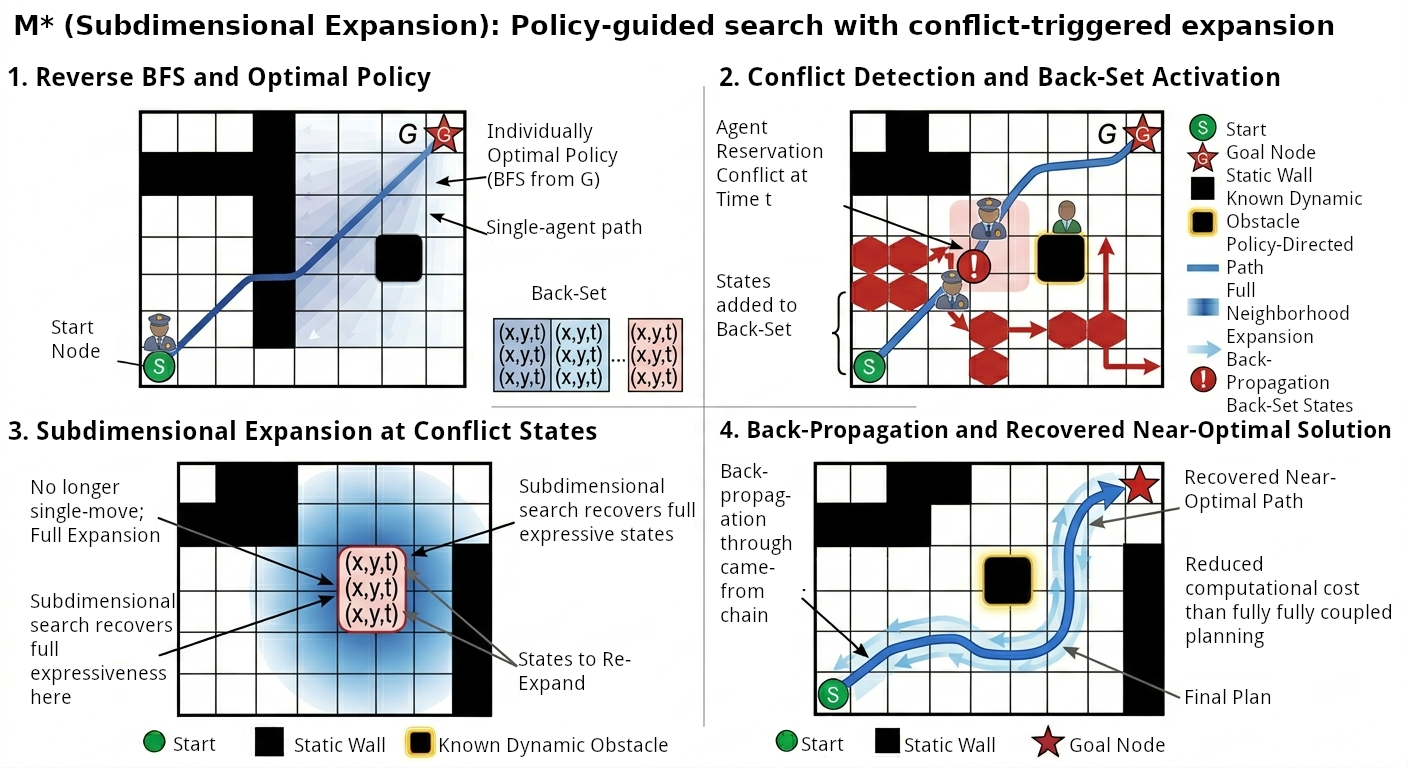}
\caption{M* visualization}
\label{fig:mstar}
\end{figure}

M* reduces the search space by initially restricting each agent to its individually optimal
policy the shortest path in the static obstacle subgraph computed by reverse BFS from the
goal. During search, agents follow their policy unless a conflict with another agent's
reservation is detected. Upon conflict, the affected states are added to a \emph{back-set},
triggering full neighborhood expansion in those regions instead of the single
policy-directed move. Back-propagation through the came-from chain ensures that previously
visited states along the path leading to the conflict point are also re-expanded, maintaining
the correctness of the search. This mechanism yields near-optimal solutions with lower
computational cost than fully coupled planning in conflict-free scenarios, while recovering
full search expressiveness wherever conflicts occur. Visualization of algorithm is presented by Fig. \ref{fig:mstar}. (\cite{MSTAR}).

\subsubsection*{Windowed Hierarchical Cooperative A* (WHCA*)}
\label{WHCA*}
\begin{figure}[H]
\centering
\includegraphics[width=\textwidth]{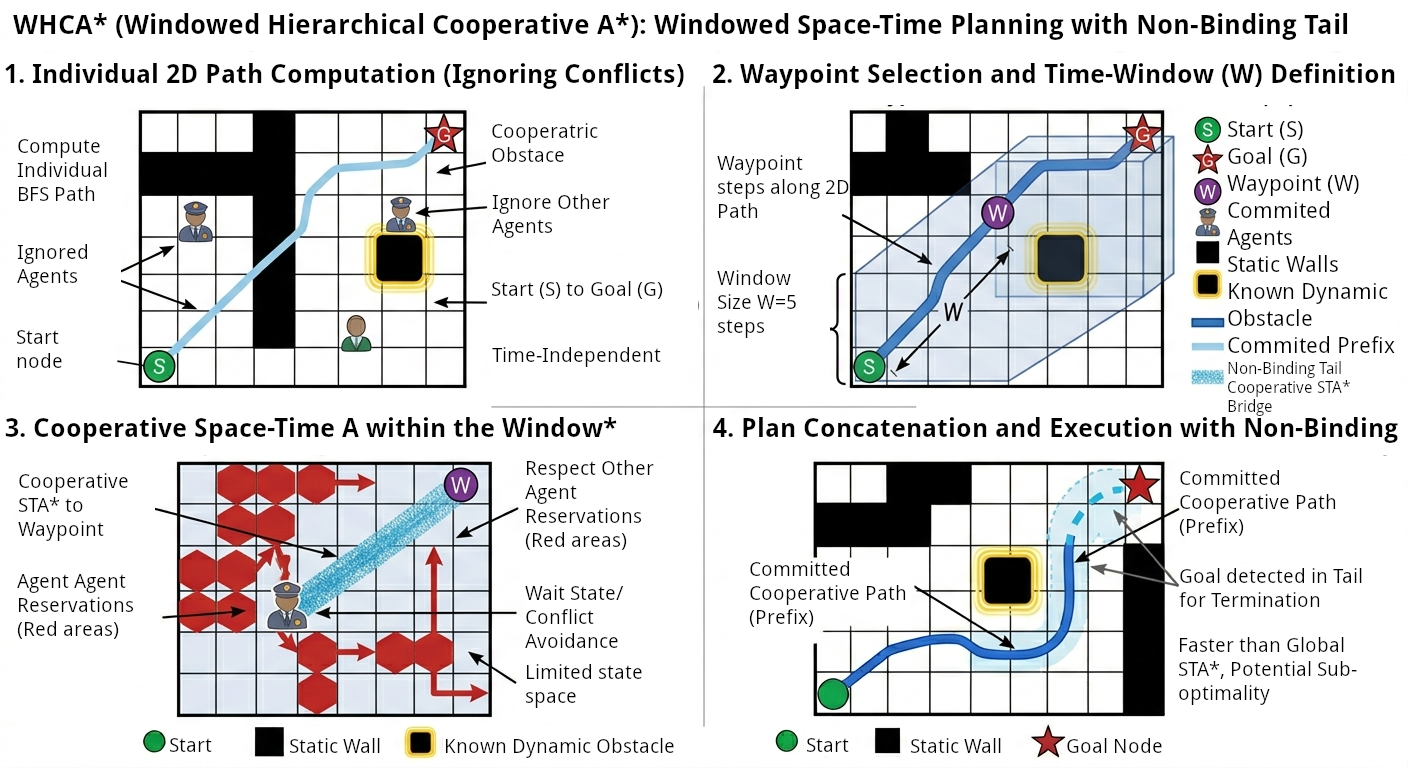}
\caption{WHCA* visualization}
\label{fig:whca}
\end{figure}

WHCA* reduces the computational cost of cooperative planning by restricting each agent's
search to a sliding time window of fixed length $W$. An individual optimal 2D path is first
computed for each agent using breadth-first search, ignoring inter-agent conflicts. A
waypoint is then selected $W$ steps along this path, and Space-Time A* is applied
cooperatively from the current position to that waypoint, respecting reservations made by
previously planned agents. The portion of the 2D path beyond the waypoint is appended as a
non-binding tail, ensuring that the planner records the goal position for termination
detection while committing only to the cooperatively planned prefix. This rolling-horizon
scheme limits the state space of each STA* invocation to $O(W \cdot |V|)$ states, making
WHCA* substantially faster than global STA* on dense multi-agent scenarios, at the cost of
potential sub-optimality. Visualization of algorithm is presented by Fig. \ref{fig:whca}. (\cite{WHCA}).

\clearpage

\clearpage

\clearpage

\clearpage
\section{Map Visualizations}
\begin{figure}
\begin{center}
    \subfigure[]
        {\label{original_path}    \includegraphics*[scale=0.2]{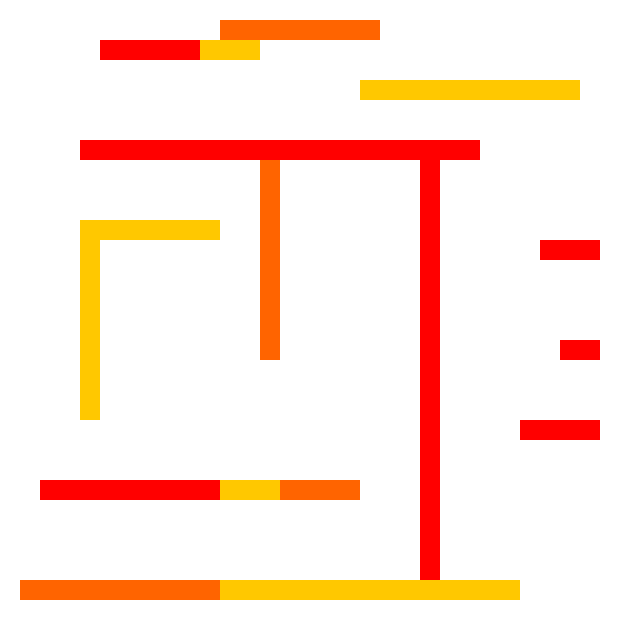}} 
      \subfigure[]
      { \label{changed_path}      \includegraphics*[scale=0.2]{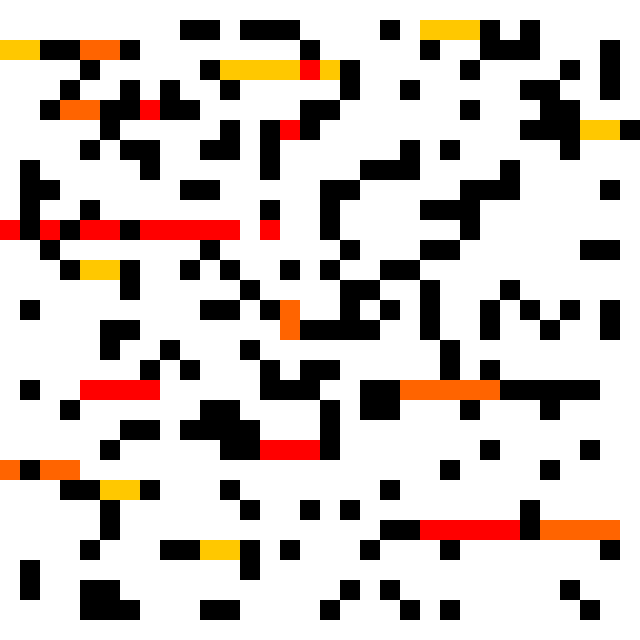}}
      \subfigure[]
      { \label{changed_path}      \includegraphics*[scale=0.2]{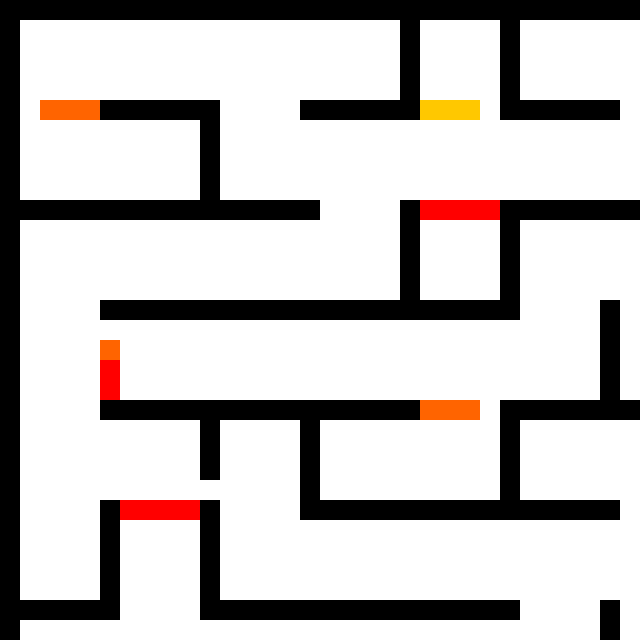}}
      \subfigure[]
      { \label{changed_path}      \includegraphics*[scale=0.2]{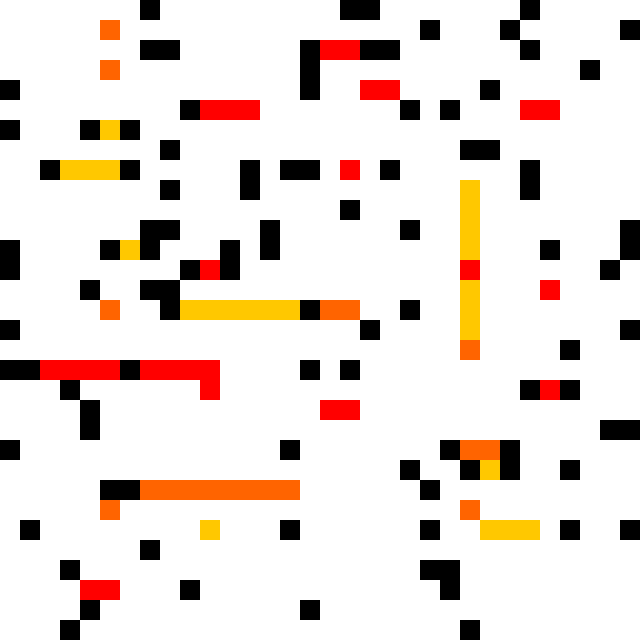}}
          \subfigure[]
        {\label{original_path}    \includegraphics*[scale=0.4]{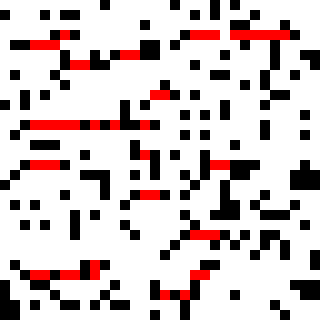}} 
      \subfigure[]
      { \label{changed_path}      \includegraphics*[scale=0.2]{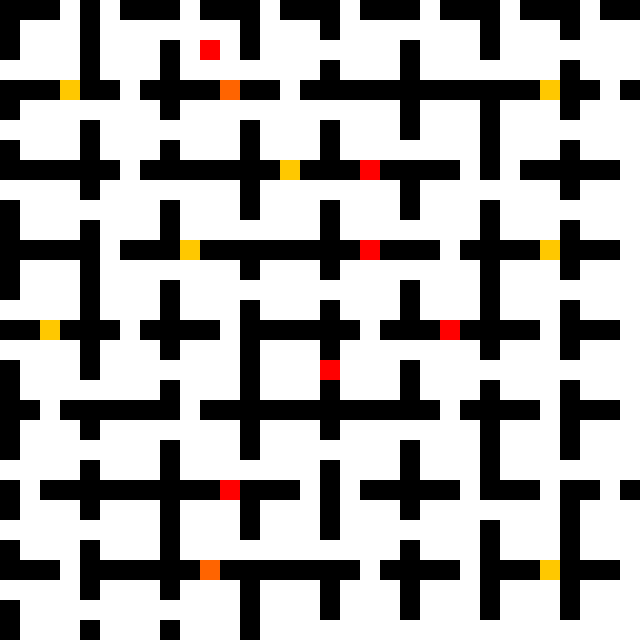}}
      \subfigure[]
      { \label{changed_path}      \includegraphics*[scale=0.10]{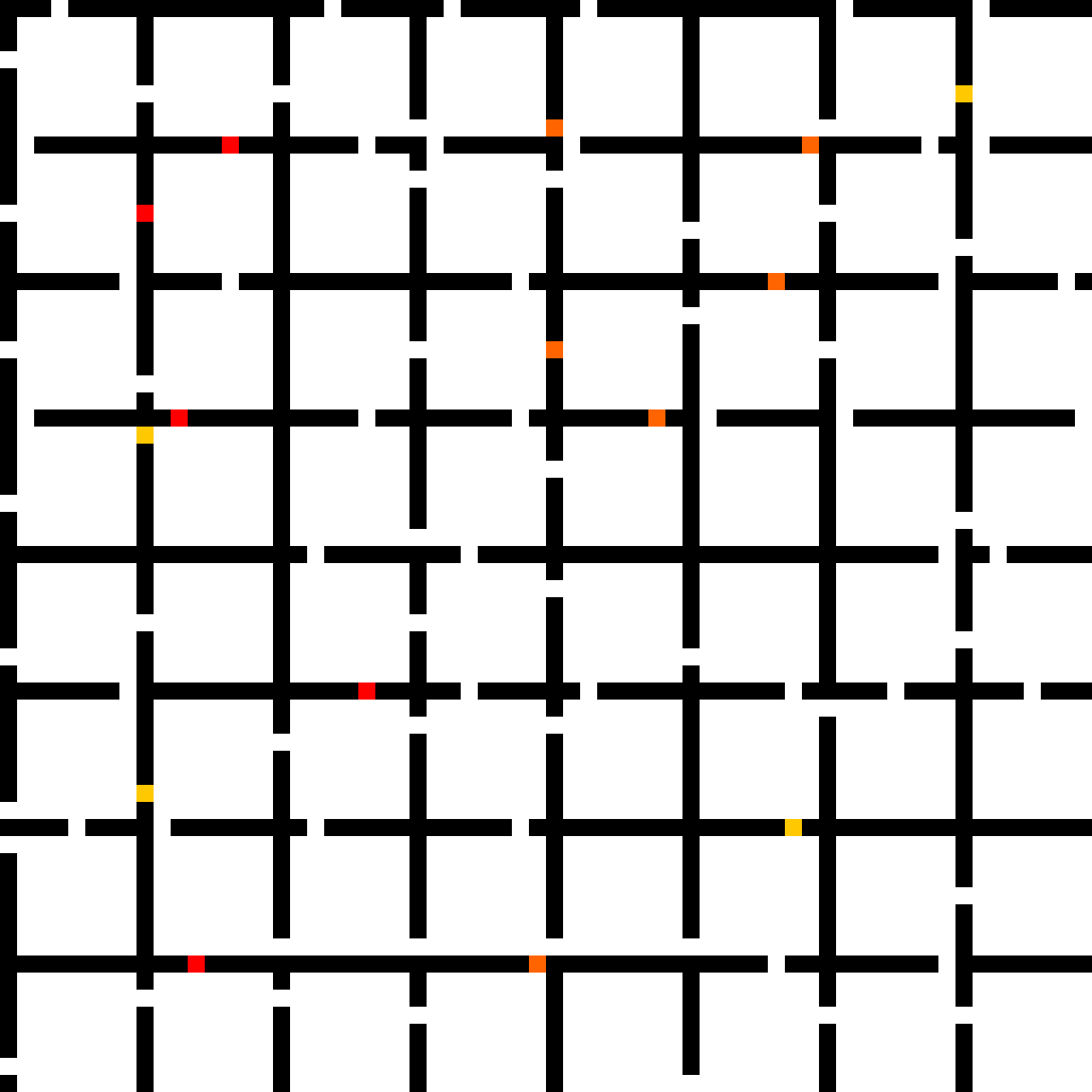}}
      \subfigure[]
      { \label{changed_path}      \includegraphics*[scale=0.10]{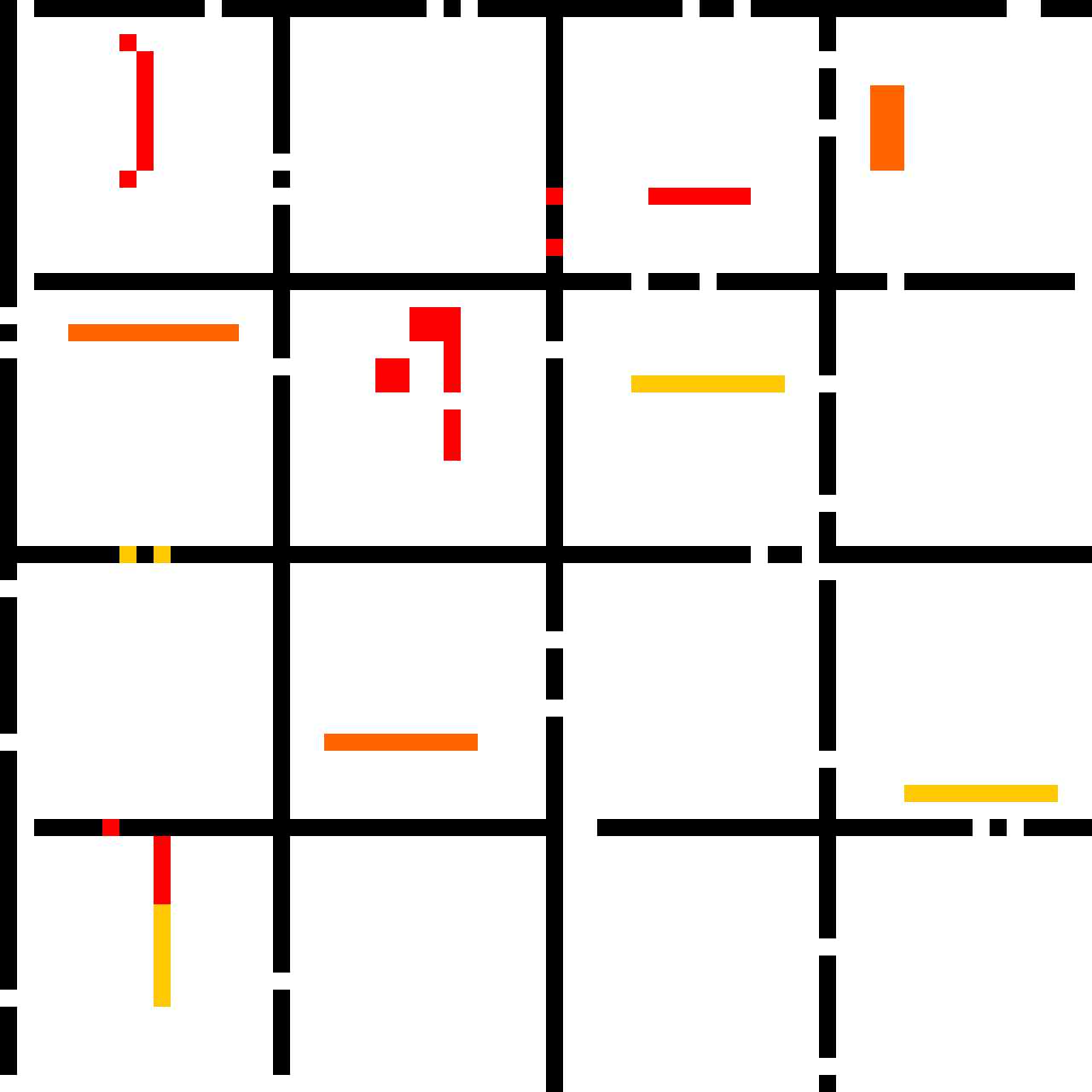}}
\caption{Visualization of maps (a) empty-32-32, (b) maze-32, (c) maze-32-32-4, (d) random-32-32-10, (e) random-32-32-20, (f) room-32-32-4, (g) room-64-64-8, (h) room-64-64-16}
  \label{maps} 
\end{center}
\end{figure}

Three colors were used to depict dynamic obstacles: red, orange and yellow. They represent variables X, Y and Z mentioned in section \ref{Dataset}. Every traversable block is white, while all static obstacles remain black. Visualization using a Python script was produced (see Fig. \ref{maps}.). The Differences between these maps are:
\begin{itemize}
    \item \textbf{empty-32-32}: A completely open grid absent of any static barriers. artificial maze is made of dynamic obstacles that change their structure depending on current randomization of activation time frame. 
    \item \textbf{maze-32}: A labyrinth layout containing tight corridors that restrict movement and frequently necessitate path recalculation (Custom Map).
    \item \textbf{maze-32-32-4}: A more intricate and compact version of the maze‐style environment, offering even greater structural confinement.
    \item \textbf{random-32-32-10} and \textbf{random-32-32-20}: Grids populated with randomly placed static blockages, resulting in irregular and less predictable traversal patterns.
    \item \textbf{room-32-32-4}, \textbf{room-64-64-8}, and \textbf{room-64-64-16}: Environments composed of rectangular chambers linked by narrow doorways. These maps vary in overall dimensions and the number of internal rooms, providing a spectrum of spatial complexity.
\end{itemize}
\clearpage
\section{Detailed Experimental Results}
\label{apendixc}

The detailed tables are presented at:
\textbf{\url{https://github.com/Miczu212/on_dynamic_multi-agent_pathfinding_methods_review_simulations_and_modifications}}
\\
\\
A comprehensive statistical analysis was performed on all simulation results. Because the performance metrics exhibit a roughly linear growth with the number of agents, presenting a single averaged value per algorithm would obscure important trends. Therefore, the results are disaggregated by agent count, with separate summaries for \(n = 1\) through \(n = 10\) (see Tables 2,3). Due to space constraints only tables for 1 and 10 agents are presented in the appendix. The full set of descriptive statistics, including standard deviations, medians, quartiles, minima, and maxima were performed. To assess whether the observed differences in solution quality are statistically reliable, we performed a Kruskal–Wallis test followed by Dunn's post‑hoc comparison with Bonferroni correction (\cite{Statistic1pt1}). The Kruskal–Wallis test was chosen because the SoC (SoC) data violated the normality assumption required for parametric ANOVA (Shapiro–Wilk test, $p < 0.01$ for most configurations (\cite{Statistic1pt2,Statistic2,Statistic3}). The analysis was conducted separately for each map and agent count, using SoC and Makespan as metrics.

The Kruskal–Wallis test rejected the null hypothesis (all algorithms perform equally) for the vast majority of map–agent combinations ($p < 0.05$). The only exceptions were configurations with a single agent (where all algorithms achieved similar SoC) and a few low‑density scenarios on large room maps (e.g., \texttt{room-64-64-16} with $n \le 3$ and $n = 10$, where the test was not significant).
The key findings can be summarized as follows:
\begin{itemize}
    \item \textbf{Solution quality}: A** achieves the lowest SoC for \(n = 1\) to \(9\), with the only exception at \(n = 10\) (see Table \ref{tab:agents10}) where M* is marginally better. Dijkstra consistently yields the highest SoC for \(n \ge 2\) due to its lack of temporal reasoning.
    \item \textbf{Computational efficiency}: Dijkstra and D* Lite are the fastest (always below 6 ms). WHCA* is roughly twice as fast as STA*. A** is two to three orders of magnitude slower, due to template generation and bridge construction.
    \item \textbf{Replanning behavior}: WHCA* and Dijkstra replan most frequently, while A** and M* replan rarely but each replan is much more expensive. A** amortizes its high offline cost by reusing templates.
    \item \textbf{Robustness}: WHCA* and STA* have the lowest failure rates (9\% and 11\% respectively (see Table \ref{tab:failure_rates_sorted})). A** and D* Lite show moderate failure rates (around 20–21\%), while M* reaches 20\% on average but peaks at 33\% for 10 agents (see Table \ref{tab:agents10}).
\end{itemize}

\label{Tables}
\clearpage
\begin{table}[p!]
\centering
\caption{Detailed results for 1 agent}
\label{tab:agents1}
\footnotesize
\begin{tabular}{c c c c c c c}
\toprule
 Algorithm & SoC & Makespan & Replans & Computation Time [ms] & Avg Replan Time [ms] & Failed Attempts [out of a 800] \\
\midrule
\begin{filecontents*}{1.csv}
A** K=9,AVG,772.00,\textbf{33.16},\textbf{33.16},196.50,0.59,25.98,28.00,"0,035"
A** K=8,AVG,769.00,\underline{33.25},\underline{33.25},174.76,0.60,29.24,31.00,"0,03875"
A** K=14,AVG,775.00,33.39,33.39,175.19,0.57,44.33,25.00,"0,03125"
A** K=7,AVG,755.00,33.76,33.76,206.89,0.78,43.86,45.00,"0,05625"
A** K=3,AVG,764.00,33.83,33.83,253.94,0.68,48.02,36.00,"0,045"
A** K=13,AVG,774.00,33.84,33.84,178.16,\textbf{0.44},47.83,26.00,"0,0325"
A** K=20,AVG,763.00,33.95,33.95,314.79,0.76,71.28,37.00,"0,04625"
A** K=19,AVG,753.00,33.98,33.98,157.22,0.69,26.41,47.00,"0,05875"
A** K=11,AVG,780.00,34.13,34.13,207.25,\underline{0.46},25.27,\underline{20.00},"0,025"
A** K=15,AVG,754.00,34.20,34.20,367.68,0.98,64.49,46.00,"0,0575"
A** K=12,AVG,777.00,34.31,34.31,197.88,0.49,42.05,23.00,"0,02875"
A** K=10,AVG,774.00,34.39,34.39,253.14,0.54,61.08,26.00,"0,0325"
A** K=16,AVG,\textbf{750.00},34.50,34.50,217.90,0.83,37.01,50.00,"0,0625"
A** K=18,AVG,759.00,34.51,34.51,188.03,0.77,47.13,41.00,"0,05125"
A** K=6,AVG,\underline{752.00},34.83,34.83,209.15,0.84,61.63,48.00,"0,06"
A** K=17,AVG,757.00,35.19,35.19,174.33,0.82,35.59,43.00,"0,05375"
A** K=5,AVG,757.00,35.58,35.58,217.81,0.96,58.25,43.00,"0,05375"
A** K=4,AVG,769.00,35.77,35.77,306.40,0.69,66.81,31.00,"0,03875"
Dijkstra,AVG,777.00,35.98,35.98,\textbf{0.49},1.58,\textbf{0.05},23.00,"0,02875"
WHCA*,AVG,782.00,36.05,36.05,2.77,1.88,0.51,\textbf{18.00},"0,0225"
D* Lite,AVG,780.00,36.74,36.74,\underline{0.52},1.62,\underline{0.06},\underline{20.00},"0,025"
M*,AVG,770.00,36.90,36.90,0.68,1.59,0.08,30.00,"0,0375"
STA*,AVG,776.00,37.11,37.11,4.13,1.85,0.89,24.00,"0,03"
\end{filecontents*}
\csvreader[no head, late after line=\\]
{1.csv}{}
{ \csvcoli & \csvcoliv & \csvcolv & \csvcolvii & \csvcolvi & \csvcolviii & \csvcolix} 
\\
\bottomrule
\end{tabular}
\end{table}

\begin{table}[htbp]
\centering
\caption{Detailed results for 10 agents}
\label{tab:agents10}
\footnotesize
\begin{tabular}{c c c c c c c}
\toprule
 Algorithm & SoC & Makespan & Replans & Computation Time [ms] & Avg Replan Time [ms] & Failed Attempts [out of a 800] \\
\midrule
\begin{filecontents*}{10.csv}
M*,AVG,533.00,\textbf{373.07},70.90,\underline{6.22},7.86,0.33,267.00,"0,33375"
A** K=10,AVG,558.00,\underline{375.03},\underline{68.63},3197.68,6.96,215.51,242.00,"0,3025"
A** K=20,AVG,527.00,376.93,\textbf{68.41},3091.57,6.53,168.29,273.00,"0,34125"
A** K=17,AVG,540.00,378.13,69.97,3068.50,\textbf{6.07},165.68,260.00,"0,325"
A** K=12,AVG,515.00,382.21,69.91,3280.41,6.94,207.99,285.00,"0,35625"
A** K=13,AVG,503.00,383.83,71.56,3413.53,7.90,218.77,297.00,"0,37125"
A** K=9,AVG,511.00,384.70,71.34,3247.01,7.46,231.57,289.00,"0,36125"
A** K=19,AVG,532.00,385.11,71.28,3582.64,6.81,243.41,268.00,"0,335"
A** K=15,AVG,533.00,385.36,70.03,3312.78,7.60,188.15,267.00,"0,33375"
D* Lite,AVG,\underline{471.00},385.88,70.51,\textbf{5.65},7.90,\textbf{0.29},329.00,"0,41125"
A** K=4,AVG,556.00,386.68,71.50,3577.46,\underline{6.09},214.96,244.00,"0,305"
WHCA*,AVG,664.00,388.62,70.75,42.30,10.26,2.67,\textbf{136.00},"0,17"
A** K=11,AVG,529.00,390.09,72.49,3771.51,7.68,204.35,271.00,"0,33875"
A** K=6,AVG,501.00,390.91,71.86,3422.09,8.35,192.43,299.00,"0,37375"
A** K=3,AVG,521.00,391.47,72.47,3359.58,7.68,201.81,279.00,"0,34875"
STA*,AVG,643.00,392.45,71.19,47.17,14.06,2.48,\underline{157.00},"0,19625"
A** K=8,AVG,530.00,393.21,72.04,3777.27,6.68,222.82,270.00,"0,3375"
A** K=16,AVG,513.00,394.85,73.65,3692.29,9.36,235.16,287.00,"0,35875"
A** K=18,AVG,513.00,395.19,73.77,3465.55,7.86,218.04,287.00,"0,35875"
A** K=5,AVG,486.00,395.29,73.29,3664.86,8.64,212.28,314.00,"0,3925"
A** K=7,AVG,488.00,403.85,74.74,4035.84,9.58,244.46,312.00,"0,39"
A** K=14,AVG,\textbf{465.00},407.87,75.34,3643.34,9.07,238.45,335.00,"0,41875"
Dijkstra,AVG,607.00,408.11,78.55,6.67,14.24,\underline{0.31},193.00,"0,24125"
\end{filecontents*}
\csvreader[no head, late after line=\\]
{10.csv}{}
{ \csvcoli & \csvcoliv & \csvcolv & \csvcolvii & \csvcolvi & \csvcolviii & \csvcolix} 
\\
\bottomrule
\end{tabular}

\end{table}

\begin{table}[htbp]
\centering
\caption{Best performing template count}
\label{tab:dunn_makespan}
\begin{tabular}{lcccccccccc}
\toprule
  \textbf{Number of agents \textit{n}:} & \textbf{1} & \textbf{2} & \textbf{3} & \textbf{4} & \textbf{5}& \textbf{6} & \textbf{7} & \textbf{8} & \textbf{9} & \textbf{10} \\
\midrule
Template count \textit{k}    & $9$ & $7$ & $12$ & $16$ & $3$ & $5$& $18$& $10$& $10$& $10$ \\
\bottomrule
\end{tabular}
\end{table}

\begin{table}[h]
\centering
\caption{Statistical Analysis for SumOfCosts}
\label{Statistical Analysis:first}
\footnotesize
\renewcommand{\arraystretch}{0.8}
\begin{tabular}{c c c c c c c c c c}
\hline
Agents & Algo & Count & Mean & Std & Median & Min & Max & Q1 & Q3 \\
\hline
\begin{filecontents*}{statistics.csv}
Agents,Algo,SumOfCosts_count,SumOfCosts_mean,SumOfCosts_std,SumOfCosts_median,SumOfCosts_min,SumOfCosts_max,SumOfCosts_q1,SumOfCosts_q3,TotalTimesteps_count,TotalTimesteps_mean,TotalTimesteps_std,TotalTimesteps_median,TotalTimesteps_min,TotalTimesteps_max,TotalTimesteps_q1,TotalTimesteps_q3,ComputationTime_ms_count,ComputationTime_ms_mean,ComputationTime_ms_std,ComputationTime_ms_median,ComputationTime_ms_min,ComputationTime_ms_max,ComputationTime_ms_q1,ComputationTime_ms_q3,TotalReplans_count,TotalReplans_mean,TotalReplans_std,TotalReplans_median,TotalReplans_min,TotalReplans_max,TotalReplans_q1,TotalReplans_q3,AvgReplanTime_ms_count,AvgReplanTime_ms_mean,AvgReplanTime_ms_std,AvgReplanTime_ms_median,AvgReplanTime_ms_min,AvgReplanTime_ms_max,AvgReplanTime_ms_q1,AvgReplanTime_ms_q3
1.00,a_star_star,13754.00,34.25,27.56,27.00,0.00,191.00,16.00,42.00,13754.00,34.25,27.56,27.00,0.00,191.00,16.00,42.00,13754.00,221.97,1315.17,8.42,0.03,59162.57,2.53,45.66,13754.00,0.69,1.73,0.00,0.00,28.00,0.00,1.00,13754.00,46.43,347.33,0.00,0.00,14694.69,0.00,0.22
2.00,a_star_star,13100.00,82.07,51.06,69.00,10.00,315.00,47.00,100.00,13100.00,50.36,31.69,41.00,5.00,200.00,29.00,62.00,13100.00,650.46,2373.08,41.72,0.51,56425.46,15.39,311.60,13100.00,1.85,2.96,1.00,0.00,40.00,0.00,3.00,13100.00,114.87,575.15,0.47,0.00,16660.58,0.00,9.40
3.00,a_star_star,12668.00,122.84,69.63,97.00,17.00,435.00,75.00,153.00,12668.00,57.43,33.37,45.00,7.00,201.00,34.00,74.00,12668.00,996.08,2937.96,65.09,0.96,58698.94,28.73,588.37,12668.00,2.35,3.23,1.00,0.00,39.00,0.00,3.00,12668.00,143.20,608.98,1.50,0.00,10427.56,0.00,15.63
4.00,a_star_star,11967.00,162.11,87.26,130.00,22.00,567.00,102.00,202.00,11967.00,61.13,34.18,48.00,6.00,201.00,37.00,77.00,11967.00,1353.54,3716.87,101.04,1.48,58174.79,43.43,891.89,11967.00,3.27,4.23,2.00,0.00,76.00,0.00,5.00,11967.00,168.06,641.88,3.37,0.00,13999.37,0.00,27.38
5.00,a_star_star,11387.00,199.83,104.79,160.00,33.00,610.00,126.00,250.00,11387.00,64.14,34.71,50.00,9.00,196.00,39.00,82.00,11387.00,1647.28,3959.80,143.11,2.94,56624.94,61.53,1236.70,11387.00,4.27,4.77,3.00,0.00,54.00,1.00,6.00,11387.00,173.05,565.98,6.00,0.00,9965.44,0.59,40.21
6.00,a_star_star,10992.00,237.03,122.83,187.00,51.00,718.00,149.00,299.00,10992.00,65.96,35.65,51.00,14.00,359.00,40.00,85.00,10992.00,2016.26,4719.94,182.09,4.93,59559.27,76.19,1567.56,10992.00,4.89,5.36,3.00,0.00,63.00,1.00,7.00,10992.00,191.08,605.12,7.65,0.00,13707.37,0.97,55.76
7.00,a_star_star,10601.00,273.82,138.77,217.00,66.00,802.00,174.00,344.00,10601.00,67.83,36.00,53.00,13.00,229.00,41.00,87.00,10601.00,2262.64,5057.17,235.46,7.09,59442.72,94.88,1889.90,10601.00,5.57,6.37,4.00,0.00,85.00,1.00,8.00,10601.00,182.59,543.30,9.76,0.00,9826.67,1.12,60.08
8.00,a_star_star,10140.00,310.79,156.17,245.00,86.00,942.00,198.00,387.00,10140.00,68.92,36.22,54.00,16.00,257.00,42.00,91.00,10140.00,2627.98,5581.45,271.70,11.30,58510.15,110.44,2167.87,10140.00,6.00,6.36,4.00,0.00,70.00,1.00,8.00,10140.00,193.36,564.79,10.58,0.00,9415.10,1.39,68.68
9.00,a_star_star,9520.00,352.77,175.03,276.00,112.00,920.00,223.00,451.00,9520.00,70.95,37.05,55.00,21.00,321.00,43.00,94.00,9520.00,3095.37,6195.00,349.51,12.45,59172.75,150.35,2928.70,9520.00,7.04,7.26,5.00,0.00,93.00,2.00,10.00,9520.00,210.62,562.01,14.22,0.00,10016.73,2.89,92.62
10.00,a_star_star,9321.00,388.65,192.09,303.00,111.00,1023.00,248.00,490.00,9321.00,71.73,37.32,56.00,21.00,254.00,44.00,96.00,9321.00,3473.40,6882.65,409.32,18.73,59931.95,168.24,3207.78,9321.00,7.59,7.83,6.00,0.00,132.00,2.00,10.00,9321.00,212.08,571.76,16.26,0.00,12948.13,3.51,96.20
1.00,d_star_lite,780.00,36.74,28.62,28.00,1.00,164.00,16.00,48.25,780.00,36.74,28.62,28.00,1.00,164.00,16.00,48.25,780.00,0.52,0.66,0.27,0.04,6.36,0.13,0.64,780.00,1.62,4.47,0.00,0.00,41.00,0.00,1.00,780.00,0.05,0.23,0.00,0.00,3.20,0.00,0.03
2.00,d_star_lite,756.00,87.20,52.79,73.00,12.00,262.00,51.00,110.25,756.00,53.37,32.35,45.00,6.00,180.00,30.00,67.00,756.00,1.30,1.30,0.87,0.10,8.14,0.47,1.64,756.00,3.76,6.76,1.00,0.00,46.00,0.00,4.00,756.00,0.13,0.36,0.05,0.00,4.76,0.00,0.11
3.00,d_star_lite,738.00,124.92,69.43,99.00,22.00,375.00,77.00,158.75,738.00,58.70,33.46,45.00,9.00,169.00,35.25,77.75,738.00,1.70,1.41,1.26,0.16,9.67,0.74,2.17,738.00,3.56,4.36,2.00,0.00,32.00,0.00,5.00,738.00,0.13,0.26,0.07,0.00,2.49,0.00,0.13
4.00,d_star_lite,715.00,173.65,92.38,139.00,30.00,490.00,108.00,216.50,715.00,65.09,35.91,52.00,9.00,181.00,39.00,87.00,715.00,2.59,2.36,1.89,0.22,24.22,1.09,3.29,715.00,5.34,6.18,3.00,0.00,36.00,1.00,8.00,715.00,0.21,0.38,0.10,0.00,2.68,0.05,0.18
5.00,d_star_lite,680.00,194.99,98.35,161.00,53.00,550.00,130.00,221.25,680.00,61.70,32.47,50.00,13.00,256.00,40.00,75.00,680.00,2.64,2.13,1.87,0.39,13.41,1.21,3.30,680.00,4.62,4.97,3.00,0.00,28.00,1.00,7.00,680.00,0.20,0.33,0.10,0.00,3.08,0.06,0.18
6.00,d_star_lite,558.00,243.75,121.28,196.00,62.00,742.00,158.25,299.75,558.00,67.56,35.29,54.00,16.00,288.00,42.00,90.75,558.00,3.86,3.26,2.64,0.51,23.54,1.67,4.97,558.00,8.21,12.48,4.00,0.00,96.00,1.00,9.00,558.00,0.25,0.34,0.13,0.00,2.16,0.10,0.21
7.00,d_star_lite,647.00,265.99,132.92,210.00,97.00,720.00,178.00,324.00,647.00,64.98,36.60,48.00,24.00,183.00,41.00,85.00,647.00,3.41,2.90,2.29,0.59,27.17,1.55,4.46,647.00,4.52,5.15,3.00,0.00,26.00,0.00,8.00,647.00,0.19,0.24,0.12,0.00,1.86,0.00,0.22
8.00,d_star_lite,547.00,315.79,160.79,244.00,124.00,797.00,203.00,381.50,547.00,69.39,38.55,53.00,23.00,311.00,43.00,89.50,547.00,5.21,4.36,3.51,0.66,25.50,2.39,6.37,547.00,10.43,8.16,9.00,0.00,35.00,4.00,15.00,547.00,0.26,0.29,0.15,0.00,2.97,0.12,0.25
9.00,d_star_lite,511.00,358.44,180.74,274.00,159.00,891.00,233.00,437.50,511.00,73.41,40.29,57.00,27.00,285.00,45.00,92.50,511.00,5.71,5.03,3.54,1.06,30.99,2.40,7.35,511.00,9.29,11.08,6.00,0.00,84.00,3.00,12.00,511.00,0.31,0.38,0.16,0.00,2.57,0.12,0.27
10.00,d_star_lite,471.00,385.87,197.71,294.00,181.00,972.00,250.00,478.50,471.00,70.51,42.17,51.00,28.00,290.00,42.00,86.00,471.00,5.65,4.45,3.91,1.23,25.58,2.30,7.79,471.00,7.90,9.43,5.00,0.00,45.00,0.00,11.00,471.00,0.29,0.39,0.17,0.00,2.04,0.00,0.30
1.00,dijkstra,777.00,35.98,27.54,29.00,1.00,163.00,17.00,46.00,777.00,35.98,27.54,29.00,1.00,163.00,17.00,46.00,777.00,0.49,0.78,0.22,0.04,8.19,0.13,0.49,777.00,1.59,4.47,0.00,0.00,37.00,0.00,1.00,777.00,0.05,0.13,0.00,0.00,0.93,0.00,0.04
2.00,dijkstra,755.00,89.42,54.89,74.00,12.00,284.00,52.00,111.00,755.00,55.54,34.39,45.00,6.00,192.00,32.00,70.50,755.00,1.19,1.39,0.72,0.13,14.38,0.39,1.45,755.00,3.67,6.53,1.00,0.00,51.00,0.00,4.00,755.00,0.12,0.18,0.06,0.00,1.24,0.00,0.14
3.00,dijkstra,751.00,136.63,71.05,115.00,26.00,392.00,86.00,168.00,751.00,63.84,32.94,54.00,10.00,172.00,39.00,83.00,751.00,1.71,1.49,1.29,0.22,15.15,0.85,2.10,751.00,4.40,4.47,3.00,0.00,35.00,1.00,6.00,751.00,0.17,0.18,0.12,0.00,1.27,0.07,0.19
4.00,dijkstra,721.00,178.74,91.99,151.00,35.00,514.00,113.00,223.00,721.00,68.36,35.92,58.00,10.00,203.00,41.00,88.00,721.00,2.23,1.70,1.77,0.27,12.69,1.04,2.86,721.00,5.63,6.17,4.00,0.00,47.00,1.00,8.00,721.00,0.19,0.22,0.13,0.00,1.75,0.07,0.19
5.00,dijkstra,714.00,212.82,102.42,183.00,54.00,583.00,142.00,253.50,714.00,69.51,34.57,58.00,16.00,221.00,44.00,86.00,714.00,2.59,2.12,1.89,0.45,16.80,1.21,3.18,714.00,5.32,5.20,4.00,0.00,31.00,1.00,8.00,714.00,0.20,0.23,0.14,0.00,2.22,0.08,0.22
6.00,dijkstra,653.00,258.66,114.92,223.00,80.00,648.00,178.00,308.00,653.00,74.72,34.91,65.00,21.00,222.00,49.00,97.00,653.00,3.90,3.27,2.69,0.70,21.72,1.71,5.00,653.00,9.99,12.88,6.00,0.00,84.00,2.00,12.00,653.00,0.23,0.23,0.16,0.00,1.33,0.12,0.23
7.00,dijkstra,683.00,289.76,144.10,235.00,85.00,793.00,195.00,325.00,683.00,73.65,37.49,60.00,19.00,191.00,47.50,87.50,683.00,3.76,3.86,2.35,0.74,30.01,1.65,3.91,683.00,5.85,5.77,4.00,0.00,39.00,2.00,8.00,683.00,0.25,0.25,0.16,0.00,2.46,0.11,0.26
8.00,dijkstra,656.00,331.99,153.47,278.50,100.00,801.00,230.75,373.50,656.00,75.47,34.73,64.00,18.00,184.00,50.00,92.25,656.00,5.24,3.96,3.66,0.82,21.64,2.38,7.20,656.00,10.84,9.18,8.00,0.00,66.00,4.00,16.00,656.00,0.28,0.24,0.19,0.00,1.39,0.15,0.27
9.00,dijkstra,648.00,375.16,168.49,313.50,104.00,899.00,258.00,431.50,648.00,78.61,34.71,69.00,18.00,216.00,52.00,98.00,648.00,5.52,4.33,3.97,0.79,36.38,2.63,7.39,648.00,10.79,11.93,7.00,0.00,75.00,4.00,14.00,648.00,0.29,0.24,0.19,0.00,2.34,0.15,0.31
10.00,dijkstra,607.00,408.11,172.44,346.00,167.00,972.00,289.50,456.00,607.00,78.55,34.23,69.00,32.00,218.00,53.50,93.00,607.00,6.67,4.02,6.06,1.47,33.19,3.45,9.03,607.00,14.25,13.11,9.00,0.00,63.00,4.00,23.00,607.00,0.31,0.24,0.23,0.00,2.05,0.19,0.31
1.00,m_star,770.00,36.90,28.63,28.00,1.00,158.00,17.00,48.00,770.00,36.90,28.63,28.00,1.00,158.00,17.00,48.00,770.00,0.68,1.19,0.34,0.12,25.73,0.18,0.75,770.00,1.59,4.05,0.00,0.00,31.00,0.00,1.00,770.00,0.08,0.19,0.00,0.00,1.82,0.00,0.12
2.00,m_star,705.00,89.12,52.13,78.00,13.00,275.00,53.00,112.00,705.00,55.29,33.06,47.00,7.00,185.00,33.00,70.00,705.00,1.44,1.31,1.07,0.23,8.81,0.53,1.78,705.00,3.45,6.00,1.00,0.00,47.00,0.00,4.00,705.00,0.18,0.28,0.14,0.00,3.00,0.00,0.17
3.00,m_star,717.00,126.59,68.84,103.00,22.00,409.00,81.00,152.00,717.00,58.85,33.82,47.00,8.00,285.00,36.00,74.00,717.00,1.79,1.45,1.37,0.40,10.79,0.79,2.26,717.00,3.16,3.99,2.00,0.00,27.00,0.00,5.00,717.00,0.20,0.28,0.15,0.00,2.69,0.00,0.19
4.00,m_star,709.00,166.79,89.72,132.00,41.00,464.00,105.00,217.00,709.00,62.40,34.99,49.00,12.00,176.00,37.00,81.00,709.00,2.39,2.04,1.66,0.49,17.30,0.94,3.19,709.00,3.49,4.66,2.00,0.00,30.00,0.00,5.00,709.00,0.24,0.35,0.16,0.00,4.54,0.00,0.23
5.00,m_star,654.00,194.17,95.24,162.50,53.00,658.00,131.25,233.75,654.00,63.46,35.57,52.50,14.00,282.00,41.00,74.00,654.00,2.62,2.01,2.08,0.73,17.71,1.25,3.34,654.00,3.87,4.03,3.00,0.00,22.00,1.00,6.00,654.00,0.24,0.27,0.18,0.00,2.75,0.14,0.23
6.00,m_star,562.00,250.07,121.35,202.00,83.00,648.00,163.00,309.75,562.00,69.26,34.77,57.00,21.00,280.00,44.00,89.00,562.00,3.94,3.01,3.01,0.77,28.86,1.83,5.16,562.00,7.04,9.59,4.00,0.00,76.00,2.00,8.00,562.00,0.29,0.29,0.20,0.00,3.06,0.17,0.31
7.00,m_star,670.00,275.51,141.27,214.00,81.00,793.00,177.25,352.75,670.00,67.59,38.01,51.00,16.00,277.00,41.00,88.00,670.00,3.74,3.11,2.55,0.78,22.70,1.58,4.96,670.00,3.96,4.61,2.00,0.00,22.00,0.00,6.00,670.00,0.26,0.33,0.20,0.00,4.29,0.00,0.29
8.00,m_star,554.00,310.29,154.37,255.50,121.00,916.00,209.00,340.00,554.00,70.79,44.32,55.00,24.00,323.00,45.00,78.00,554.00,5.59,4.52,4.28,1.21,30.16,2.74,6.70,554.00,9.54,7.99,7.00,0.00,44.00,3.00,14.00,554.00,0.33,0.29,0.22,0.00,2.08,0.20,0.28
9.00,m_star,545.00,342.21,167.45,275.00,128.00,977.00,232.00,392.00,545.00,70.81,41.08,56.00,23.00,313.00,46.00,79.00,545.00,5.69,4.35,4.08,1.15,36.88,2.73,7.64,545.00,8.08,8.17,6.00,0.00,50.00,3.00,10.00,545.00,0.37,0.36,0.23,0.00,2.88,0.21,0.34
10.00,m_star,533.00,373.07,181.45,295.00,153.00,993.00,251.00,433.00,533.00,70.90,41.79,52.00,27.00,295.00,43.00,86.00,533.00,6.22,5.76,4.07,1.47,64.92,2.29,8.63,533.00,7.86,9.97,5.00,0.00,83.00,1.00,11.00,533.00,0.33,0.36,0.24,0.00,3.33,0.20,0.35
1.00,space_time_astar,776.00,37.11,28.75,30.00,1.00,163.00,17.00,47.00,776.00,37.11,28.75,30.00,1.00,163.00,17.00,47.00,776.00,4.13,28.29,0.17,0.02,588.60,0.07,0.65,776.00,1.85,4.48,0.00,0.00,40.00,0.00,2.00,776.00,0.89,7.51,0.00,0.00,156.82,0.00,0.02
2.00,space_time_astar,777.00,85.15,50.22,73.00,13.00,278.00,50.00,105.00,777.00,52.16,31.30,44.00,7.00,183.00,31.00,63.00,777.00,6.86,31.19,0.81,0.07,537.36,0.29,2.74,777.00,3.30,4.46,2.00,0.00,30.00,0.00,5.00,777.00,1.10,6.09,0.03,0.00,90.14,0.00,0.12
3.00,space_time_astar,753.00,126.44,67.43,103.00,23.00,370.00,79.00,155.00,753.00,58.29,32.23,46.00,9.00,176.00,35.00,75.00,753.00,23.13,355.39,1.21,0.11,9666.99,0.55,4.91,753.00,4.19,5.52,2.00,0.00,34.00,0.00,6.00,753.00,2.30,23.39,0.05,0.00,603.30,0.00,0.16
4.00,space_time_astar,701.00,161.51,82.76,132.00,29.00,465.00,105.00,191.00,701.00,60.44,32.47,49.00,11.00,181.00,38.00,75.00,701.00,15.71,64.76,1.26,0.15,861.23,0.68,6.81,701.00,3.80,4.77,2.00,0.00,34.00,0.00,5.00,701.00,1.55,5.15,0.06,0.00,49.72,0.00,0.24
5.00,space_time_astar,709.00,195.44,97.63,161.00,54.00,552.00,132.00,226.00,709.00,63.00,33.58,49.00,17.00,189.00,40.00,77.00,709.00,23.48,167.35,2.15,0.26,3839.71,1.14,8.25,709.00,6.61,6.69,5.00,0.00,47.00,2.00,10.00,709.00,1.94,15.08,0.09,0.00,383.80,0.05,0.28
6.00,space_time_astar,712.00,233.36,114.41,188.50,71.00,628.00,154.00,284.75,712.00,64.66,34.27,53.00,17.00,196.00,39.00,80.00,712.00,14.07,32.35,2.35,0.30,349.66,1.05,12.00,712.00,5.72,6.94,3.00,0.00,36.00,1.00,8.00,712.00,1.49,4.90,0.09,0.00,55.95,0.05,0.35
7.00,space_time_astar,705.00,270.20,132.71,217.00,90.00,785.00,178.00,332.00,705.00,66.23,35.40,53.00,21.00,298.00,42.00,81.00,705.00,25.67,133.40,2.49,0.42,2607.73,1.24,16.15,705.00,5.25,5.49,4.00,0.00,41.00,1.00,8.00,705.00,1.88,7.22,0.11,0.00,79.24,0.06,0.60
8.00,space_time_astar,672.00,303.51,145.83,243.00,117.00,787.00,202.00,367.00,672.00,67.28,34.18,53.00,19.00,196.00,42.00,84.00,672.00,56.59,460.26,2.92,0.45,8848.37,1.64,16.50,672.00,6.48,6.48,5.00,0.00,54.00,2.00,9.00,672.00,3.36,20.58,0.12,0.00,298.74,0.08,0.57
9.00,space_time_astar,663.00,357.47,165.35,285.00,144.00,860.00,242.00,418.00,663.00,71.80,35.89,58.00,26.00,269.00,45.00,87.00,663.00,33.64,85.55,6.92,0.77,1232.64,2.41,24.40,663.00,12.08,14.49,8.00,0.00,84.00,3.00,15.00,663.00,2.54,9.16,0.17,0.00,131.84,0.11,1.00
10.00,space_time_astar,643.00,392.45,187.44,317.00,138.00,935.00,260.50,448.00,643.00,71.19,36.98,55.00,27.00,333.00,46.00,87.00,643.00,47.17,258.41,8.12,0.58,4599.08,2.19,27.54,643.00,14.06,23.66,6.00,0.00,148.00,2.00,13.00,643.00,2.48,11.38,0.21,0.00,207.68,0.11,1.15
1.00,whca_star,782.00,36.05,27.21,29.00,1.00,182.00,18.00,45.00,782.00,36.05,27.21,29.00,1.00,182.00,18.00,45.00,782.00,2.77,32.55,0.17,0.02,743.93,0.11,0.37,782.00,1.88,4.94,0.00,0.00,41.00,0.00,1.00,782.00,0.51,7.76,0.00,0.00,210.43,0.00,0.02
2.00,whca_star,753.00,88.10,49.79,76.00,12.00,303.00,55.00,105.00,753.00,54.54,31.96,47.00,6.00,189.00,33.00,66.00,753.00,5.53,31.27,0.65,0.10,450.61,0.35,1.15,753.00,3.84,5.23,2.00,0.00,36.00,0.00,5.00,753.00,1.27,8.04,0.06,0.00,121.38,0.00,0.12
3.00,whca_star,757.00,127.63,69.46,104.00,23.00,384.00,81.00,154.00,757.00,59.30,33.85,47.00,9.00,175.00,36.00,75.00,757.00,16.55,232.57,0.91,0.15,6290.32,0.51,1.60,757.00,4.03,5.22,2.00,0.00,34.00,0.00,5.00,757.00,1.98,19.38,0.08,0.00,483.74,0.00,0.14
4.00,whca_star,737.00,165.85,87.85,131.00,27.00,477.00,106.00,203.00,737.00,60.92,33.35,49.00,9.00,180.00,37.00,76.00,737.00,19.03,287.36,1.19,0.21,7692.67,0.67,2.03,737.00,4.53,5.54,3.00,0.00,35.00,0.00,7.00,737.00,1.51,13.49,0.08,0.00,320.46,0.00,0.17
5.00,whca_star,725.00,203.97,105.50,164.00,51.00,585.00,132.00,242.00,725.00,65.24,35.72,52.00,14.00,190.00,41.00,79.00,725.00,15.11,64.51,1.50,0.33,689.58,0.94,2.49,725.00,5.51,6.06,4.00,0.00,53.00,1.00,8.00,725.00,1.78,7.93,0.10,0.00,85.90,0.06,0.18
6.00,whca_star,712.00,240.03,122.32,191.00,61.00,666.00,155.00,291.25,712.00,66.65,36.11,53.00,15.00,195.00,40.00,86.00,712.00,15.83,83.51,1.76,0.41,1643.88,1.03,3.15,712.00,4.95,5.90,3.00,0.00,57.00,1.00,7.00,712.00,1.61,6.75,0.11,0.00,82.11,0.05,0.29
7.00,whca_star,749.00,278.57,140.68,222.00,82.00,747.00,180.00,347.00,749.00,67.20,36.64,51.00,16.00,178.00,40.00,90.00,749.00,13.01,63.78,2.19,0.50,700.93,1.27,3.69,749.00,4.92,5.39,3.00,0.00,36.00,0.00,8.00,749.00,1.40,9.19,0.14,0.00,200.92,0.00,0.30
8.00,whca_star,684.00,316.73,156.85,246.50,126.00,859.00,208.75,388.00,684.00,69.59,36.18,54.00,25.00,182.00,44.00,86.00,684.00,20.02,90.21,3.19,0.65,1263.65,1.89,5.37,684.00,9.75,10.60,6.00,0.00,57.00,2.00,13.00,684.00,1.53,6.25,0.16,0.00,63.78,0.11,0.32
9.00,whca_star,679.00,354.06,174.74,275.00,148.00,942.00,232.00,439.50,679.00,70.65,37.01,53.00,30.00,264.00,45.00,85.00,679.00,54.47,466.17,3.66,0.95,9139.11,2.25,6.39,679.00,9.09,7.74,7.00,0.00,60.00,3.00,13.50,679.00,3.32,21.31,0.17,0.00,415.28,0.12,0.44
10.00,whca_star,664.00,388.62,191.87,301.50,149.00,1063.00,256.75,466.75,664.00,70.75,38.40,53.00,22.00,193.00,43.00,84.00,664.00,42.30,328.09,3.93,1.03,7907.14,2.16,9.14,664.00,10.26,15.18,7.00,0.00,138.00,2.00,11.00,664.00,2.67,17.56,0.20,0.00,395.22,0.11,0.48
\end{filecontents*}
\csvreader[head to column names]{statistics.csv}{}{
\csvcoli & \csvcolii & \csvcoliii & \csvcoliv & \csvcolv & \csvcolvi & \csvcolvii & \csvcolviii & \csvcolix & \csvcolx \\
}
\end{tabular}
\end{table}
\clearpage
\begin{table}[htbp]
\centering
\caption{Number of statistically significant pairwise differences (Dunn test, Bonferroni-corrected p $<$ 0.05) for SoC.}
\label{tab:dunn_soc}
\begin{tabular}{lcccccc}
\toprule
 & \textbf{STA*} & \textbf{WHCA*} & \textbf{M*} & \textbf{A**} & \textbf{D* Lite} & \textbf{Dijkstra} \\
\midrule
STA*     & -- & 2 & 15 & 8 & 42 & 48 \\
WHCA*    & 2 & -- & 12 & 6 & 39 & 44 \\
M*       & 15 & 12 & -- & 18 & 38 & 40 \\
A**      & 8 & 6 & 18 & -- & 51 & 53 \\
D* Lite  & 42 & 39 & 38 & 51 & -- & 26 \\
Dijkstra & 48 & 44 & 40 & 53 & 26 & -- \\
\bottomrule
\end{tabular}
\end{table}

\begin{table}[htbp]
\centering
\caption{Number of statistically significant pairwise differences (Dunn test, Bonferroni-corrected p $<$ 0.05) for Makespan.}
\label{tab:dunn_makespan}
\begin{tabular}{lcccccc}
\toprule
 & \textbf{STA*} & \textbf{WHCA*} & \textbf{M*} & \textbf{A**} & \textbf{D* Lite} & \textbf{Dijkstra} \\
\midrule
STA*     & -- & 3 & 12 & 7 & 35 & 40 \\
WHCA*    & 3 & -- & 10 & 5 & 32 & 37 \\
M*       & 12 & 10 & -- & 14 & 30 & 33 \\
A**      & 7 & 5 & 14 & -- & 42 & 45 \\
D* Lite  & 35 & 32 & 30 & 42 & -- & 21 \\
Dijkstra & 40 & 37 & 33 & 45 & 21 & -- \\
\bottomrule
\end{tabular}
\end{table}
\clearpage
\bibliographystyle{spbasic}
\bibliography{cas-refs}

\end{document}